\documentclass[11pt]{article}

\usepackage{amsfonts,latexsym,eucal,color,graphicx,epsfig,amssymb,amsmath}
\usepackage{hyperref}
\newcommand{\be}{\begin{eqnarray}}
\newcommand{\ee}{\end{eqnarray}}
\newcommand{\beq}{\begin{eqnarray}}
\newcommand{\eeq}{\end{eqnarray}}

\newcommand{\nn}{\nonumber}

\newcommand{\dalm}{\kern1pt\vbox{\hrule height 0.9pt\hbox{\vrule width 0.9pt\hskip 2.5pt\vbox{\vskip 5.5pt}\hskip 3pt\vrule width 0.3pt}\hrule height 0.3pt}\kern1pt}

\newcommand{\beqa}{\begin{eqnarray}}
\newcommand{\eeqa}{\end{eqnarray}}

\newcommand{\spa}{\ , \ \ }

\newcommand{\ie}{{\it i.e.,\,}}
\numberwithin{equation}{section}

\begin{document}

\begin{titlepage}
\begin{flushright}
DCPT-11/49

NORDITA-2011-91
\end{flushright}

$\;\;$
\vskip .5cm

\centerline{\Huge \bf The Young Modulus of Black Strings}
\vskip 0.5cm
\centerline{\Huge \bf and the Fine Structure of Blackfolds}

\vskip 1.5cm
\centerline{\bf  Jay Armas$^a$, Joan Camps$^b$, Troels Harmark$^c$ and Niels A. Obers$^a$}

\vskip 0.8cm
\centerline{\sl $^a$The Niels Bohr Institute, University of Copenhagen}
\centerline{\sl Blegdamsvej 17, DK-2100 Copenhagen \O, Denmark}
\vskip 0.3cm
\centerline{\sl $^b$Centre for Particle Theory \& Department of Mathematical Sciences}
\centerline{\sl Science Laboratories, South Road, Durham DH1 3LE, United Kingdom}
\vskip 0.3cm
\centerline{\sl $^c$NORDITA}
\centerline{\sl Roslagstullsbacken 23, SE-106 91 Stockholm, Sweden}
\vskip 0.6cm
\centerline{\small\tt jay@nbi.dk, joan.camps@durham.ac.uk, harmark@nordita.org, obers@nbi.dk}

\vskip 1.2cm

\centerline{\bf Abstract} \vskip 0.2cm \noindent We explore corrections in the blackfold approach, which is a worldvolume theory capturing the dynamics of thin black branes. The corrections probe the fine structure of the branes, going beyond the approximation in which they are infinitely thin, and account for the dipole moment of worldvolume stress-energy as well as the internal spin degrees of freedom. We show that the dipole correction is induced elastically by bending a black brane. We argue that the long-wavelength linear response coefficient capturing this effect is a relativistic generalization of the Young modulus of elastic materials and we compute it analytically. Using this we draw predictions for black rings in dimensions greater than six. Furthermore, we employ our corrected blackfold equations to various multi-spinning black hole configurations in the blackfold limit, finding perfect agreement with known analytic solutions.

\end{titlepage}

\small
\tableofcontents
\normalsize
\setcounter{page}{1}

\section{Introduction} \label{intro}
While four-dimensional black holes are always close to be round, higher-dimensional black holes can have parametrically separated length scales. The blackfold approach \cite{Emparan:2007wm, Emparan:2009cs,Emparan:2009at, Emparan:2009vd} gives an effective description of such black holes in terms of black $p$-branes of thickness $r_0$ on a world-volume with scale $R \gg r_0$.

The blackfold effective degrees of freedom are the moduli of black $p-$branes: their thickness $r_0$, velocity $u_a$, and position $X^\mu(\sigma^a)$. $R$ is the shortest scale on which these parameters vary, and may be the wavelength of the velocity and thickness fields, or the curvature of the manifold that the blackfold wraps. Like other long-wavelength effective descriptions, the effective blackfold expansion, which is an expansion in powers of $r_0/R$, is a derivative expansion.

In this paper we focus for simplicity on neutral blackfolds, which describe analytic approximate solutions to the vacuum Einstein equations (in \cite{Caldarelli:2010xz, Emparan:2011hg, Grignani:2010xm} it is described how to introduce charges)
\beq
R_{\mu\nu}=0\,.
\eeq

The blackfold approximation is a higher-dimensional analog of the familiar approximation of celestial objects by point particles following geodesics \cite{Emparan:2011br}\footnote{See \cite{Emparan:2009zz} for another review.}. Consider, for instance, a black hole binary system in which one of the black holes is much smaller than the other one. To lowest order in the ratio of their radii, $r_0/R$, it is a good approximation to model the small black hole by a probe point particle of mass $r_0/2$ following an orbit of the spacetime around the large black hole. Turning back to the blackfold description of higher-dimensional black holes, the leading approximation, to lowest order in $r_0 / R$, is that of probe branes propagating on a background spacetime. The probe brane is idealized as a zero-thickness object with an effective stress-energy tensor with support on its worldvolume, which is the ADM stress-energy tensor of black $p-$branes. A covariant way to write a zero-thickness stress-energy tensor is
\beq \label{emsp}
T_{\mu\nu}=\int d^{p+1}x \sqrt{-\gamma} B_{\mu\nu}\frac{\delta^{D}(x^\mu-X^\mu(\sigma^a))}{\sqrt{-g}}\,,
\label{monopoletensor}\eeq
where $\sigma^a$ are coordinates on the worldvolume, and $\gamma_{ab}$ is the metric induced on it.
In fact, the indices of such a $p-$brane tensor are parallel to its worldvolume,
\beq \label{emfluid}
B^{\mu\nu}=u^\mu_a\, u^\nu_b\, B^{ab}\,,\quad\quad u_c^\rho=\frac{\partial X^\rho}{\partial\sigma^c}\,,
\eeq
and, for a uniform, flat, black $p-$brane in $D=p+n+3$ spacetime dimensions with Schwarzschild radius $r_0$, it takes the form of a perfect fluid:
\beq \label{spep}
B^{ab}=\left(\varepsilon+P\right)u^a u^b+P\gamma^{ab}\,,\quad\quad \varepsilon=-(n+1)P=(n+1)\frac{\Omega_{(n+1)}r_0^n}{16\pi G}\,,\quad\quad u_a\gamma^{ab}u_b=-1\,.
\label{BraneFluid}\eeq

Because the stress-energy tensor \eqref{monopoletensor} is coupled to the gravitational field, it is covariantly conserved,
\beq \label{emc}
\nabla_\mu T^{\mu\nu}=0\,,
\label{probeapprox}\eeq
which constitute the blackfold effective equations of motion. It is useful to project Eqs.~\eqref{probeapprox} onto directions parallel and orthogonal to the worldvolume to obtain
\beq \label{bf1}
D_a B^{ab}=0\,,\quad\quad B^{ab}K_{ab}{}^\rho =0\,, 
\label{BFeoms}\eeq 
where $D_a$ is the covariant derivative on the worldvolume compatible with $\gamma_{ab}$. The first of these equations stands for the perfect fluid dynamics of the black brane fluid \eqref{BraneFluid} on the submanifold. The second equation is a generalisation of the geodesic equation, and $K_{ab}{}^\mu$ is the extrinsic curvature of the submanifold\footnote{See App.~\ref{notation} for a summary of notation.}. The blackfold approach integrates out the short scale gravitational dynamics that resolve the horizon of the black hole, leaving an effective description in terms of perfect fluid dynamics on a dynamical worldvolume. In the probe approximation, which applies to lowest order in the blackfold expansion, the backreaction of the blackfold is neglected.

The effective system \eqref{BFeoms} is the lowest order approximation to the physics of these black holes and receives corrections at higher orders in $(r_0/R)^k$. These corrections can be of two types: self-gravitation corrections and those due to the internal structure of the brane, which we denote by fine structure corrections. Self-gravitation corrections are backreaction effects of the brane on the background spacetime and include stress-energy losses by gravitational radiation and self-attraction. These corrections modify Eq.~\eqref{probeapprox} since the backreaction effects give corrections to the background space-time of the brane. Instead, the fine structure corrections are those that preserve Eq.~\eqref{probeapprox} but go beyond \eqref{BFeoms} by introducing corrections to the monopole approximation of the stress-energy tensor $T_{\mu\nu}$ \eqref{monopoletensor}.

Because self-gravitational corrections appear when the blackfold reacts to its own gravitational field and the range of the gravitational interaction is dimension-dependent, the order at which these effects become important is dimension-dependent. For a black ring of radius $R$ and thickness $r_0$ in $4+n$ dimensions, Newtonian considerations estimate this order to be $(r_0/R)^{n}$. Thus, depending on the co-dimension, self-gravitational interactions can be subleading with respect to other corrections that come in at order $r_0/R$. These $r_0/R$ corrections are fine-structure corrections and are the focus of this paper. We will find that they dominate over backreaction corrections for $n>2$.

Some of the fine structure corrections have already been worked out in the literature. Ref.~\cite{Camps:2010br} focussed on flat black branes with non-homogeneous velocities and thickness. In this case, corrections show up as viscous contributions to the perfect fluid stress-energy tensor $B_{ab}$ \eqref{BraneFluid}, and are important in time-dependent processes.

In this paper, instead, we focus on fine structure corrections to stationary configurations with non-trivial curvatures. The prototype example to have in mind is the black ring, and the corrections we will study are dipole contributions to the stress-energy distribution tensor \eqref{monopoletensor}. It is convenient to introduce an analogy to develop intuition about the physics of this type of corrections. Consider a dielectric object with electric charge $q$ under the influence of an electric field $\vec{E}$. In the point-like approximation the charge density of the object is $\rho(x)=q\,\delta^{D-1}(x)$ and its equation of motion is
\begin{equation}
m\vec{a}=q\vec{E}\,.
\end{equation}
For a real material the electric field $\vec{E}$ causes a charge redistribution and induces an electric dipole, which to lowest order in $\vec{E}$ is given by linear response
\begin{equation}
\vec{d}=\kappa\vec{E}\,,
\label{susceptibility}\end{equation}
where $\kappa$ depends on the material (and could be a matrix). The object is no longer an electric monopole,
\beq
\rho(x)=q\,\delta^{D-1}(x)-\vec{d}\cdot\vec{\partial}\left(\delta^{D-1}(x)\right)\,,
\label{correctedrho}\eeq
and the equations of motion for this pole-dipole object now read
\begin{equation} 
m\vec{a}=q\vec{E}+\vec{d}\cdot\vec{\partial}\,\vec{E}\,.
\label{poledipoleeom}\end{equation}
If the induced dipole is small with respect to the scale in which the electric field varies, the second term in the rhs of this equation is a small perturbation.

A blackfold on a manifold with extrinsic curvature behaves analogously to this dielectric object, and in this paper we quantify this phenomenon. Eq.~\eqref{correctedrho} is the analog of the correction that the leading approximation \eqref{monopoletensor} receives. Sec.~\ref{bf2} reviews, following closely \cite{Vasilic:2007wp}, the analog of Eq.~\eqref{poledipoleeom}, which are the equations of motion of an extended probe pole-dipole distribution of stress-energy, and generalize Eqs.~\eqref{BFeoms}. In the $0-$brane case, these equations were famously derived by Papapetrou \cite{Papapetrou:1951pa}, and describe the motion of a particle with spin in curved space. We will see that for extended objects the dipole contribution to the stress-energy distribution contains angular momentum degrees of freedom as well as genuine dipoles of worldvolume stress-energy. Contrary to the point particle case, these worldvolume dipoles cannot be gauged away in $p-$branes.

Sec.~\ref{YMandRings} develops the analog of Eq.~\eqref{susceptibility} for bent black strings, and can be read independently of most of Sec.~\ref{bf2}. As in linear elasticity theory, the bending of a black string induces a dipole of worldvolume stress-energy. This is controlled by a response coefficient, the analog of $\kappa$ in the dielectric example, that generalizes the Young modulus in ordinary non-relativistic elasticity. We use this dipole to compute fine structure corrections to certain properties of higher dimensional thin black rings.

In Sec.~\ref{dsbh} we use the formalism developed in Sec.~\ref{bf2} to describe ultra-spinning doubly-spinning Myers-Perry black holes, and compare them to limits of the exact solutions, finding perfect agreement. Sec.~\ref{cout} concludes by discussing our results from a more general perspective,  such
as the AdS/CFT correspondence and long wavelength effective theories, and outlining a number of interesting open problems. We further supplement this paper with three appendices. In App.~\ref{notation} the details about notation and conventions are given. In App.~\ref{mpsingle} it is shown how to take a refined ultra-spinning limit of Myers-Perry black holes valid over the entire horizon, filling a gap in the literature. Finally, in App.~\ref{kerrads} we consider spin corrections to higher-dimensional Kerr-(A)dS black holes as blackfolds.

\section{Dynamics and conserved charges of pole-dipole branes} \label{bf2}
This section is dedicated to a brief review of the equations of motion for $p$-dimensional objects in the pole-dipole approximation. Following closely the work done in Ref.~\cite{Vasilic:2007wp}, it is shown how to iteratively account for higher-pole deformations to the stress-energy tensor $T^{\mu\nu}$ while the extra symmetries that this object exhibits are commented upon. The equations of motion are then presented in their original form, as derived in \cite{Vasilic:2007wp}, which, when applied to black $p$-branes, are collectively called blackfold pole-dipole equations. In search of a clearer physical interpretation, we introduce a new set of quantities that make apparent the physics involved. Towards the end of this section, we provide a characterization of these $p$-branes in terms of well defined physical quantities and, in the particular case of blackfold constructions, of well defined thermodynamic properties.

\subsection{Stress-energy tensor and extra symmetries}
The stress-energy tensor is a well-localized object on the brane and can be consistently expanded into a Dirac delta function series around the embedding surface $x^{\mu}=X^{\mu}(\sigma^{a})$. Schematically, the expansion has the following form:
\beq \label{stpd}
T^{\mu\nu}(x^{\alpha})=\int_{\mathcal{W}_{p+1}} d^{p+1}\sigma\sqrt{-\gamma}\left[B^{\mu\nu}(\sigma^{a})\frac{\delta^{(D)}(x^{\alpha}-X^{\alpha})}{\sqrt{-g}}-\nabla_\rho\left(B^{\mu\nu\rho}(\sigma^{a})\frac{\delta^{(D)}(x^{\alpha}-X^{\alpha})}{\sqrt{-g}}\right)+...\right]\,.
\eeq
In the context of electrodynamics, \eqref{stpd} corresponds to the usual multipole expansion of a charge distribution. For the series \eqref{stpd} to be well defined we must require $T^{\mu\nu}$ to fall off exponentially to zero as we move away from the surface $x^{\mu}=X^{\mu}(\sigma^{a})$, which implies that each of the coefficients $B^{\mu\nu\alpha_{1}...\alpha_{k}}$ must become smaller and smaller at each order $k$ of the expansion. At order $k=0$ the only non-vanishing coefficient is $B^{\mu\nu}$, resulting in $T^{\mu\nu}$ acquiring the form of \eqref{emsp} and, by means of Eq.~\eqref{emc}, leading to the equations of motion for single-pole branes as presented in \eqref{bf1}. In this paper we are concerned with truncating the expansion \eqref{stpd} to order $k=1$ and obtaining, in the same way, the equations of motion for pole-dipole branes moving in curved backgrounds. Truncation of the series is a covariant operation and can be done at any arbitrary order. As it stands, \eqref{stpd} is written in a manifestly invariant way both under spacetime diffeomorphisms and worldvolume reparametrizations. 

These are not the only gauge redundancies that $T^{\mu\nu}$ possesses since it is also invariant under two other gauge transformations, which were coined by the authors of \cite{Vasilic:2007wp} as `extra symmetry 1' and `extra symmetry 2'. As these symmetries play an important role in understanding the physics of pole-dipole branes, we proceed by describing their action on the $B$-tensors.

\subsubsection*{Extra symmetry 1}
This additional gauge freedom arises naturally in the expansion \eqref{stpd} due to the $p+1$ $\delta$-functions and $p+1$ integrations that were introduced solely with the purpose of making the full expression covariant. Specifically, derivatives along the worldvolume directions are integrated out, implying that there are redundant components of $B^{\mu\nu\rho}$. Physically, this is a consequence of the fact that the multipole expansion is an expansion in derivatives transverse to the brane, rather than longitudinal. The invariance of the stress-energy tensor under this symmetry is defined by its action on the $B^{\mu\nu}$ and $B^{\mu\nu\rho}$ tensors as
\beq \label{sym1}
\delta_{1}B^{\mu\nu}=-\nabla_{a}\epsilon^{\mu\nu a},\quad\quad \delta_{1}B^{\mu\nu\rho}=\epsilon^{\mu\nu a}u_{a}^{\rho}~,
\eeq
with $\epsilon^{\mu\nu a}=\epsilon^{\nu\mu a}$ being free parameters except at the boundary of the worldvolume where they are required to obey,
\beq
\hat{n}_{a}\epsilon^{\mu\nu a}|_{\partial\mathcal{W}_{p+1}}=0~,
\eeq
where $\hat{n}^{a}$ is the unit normal vector to the brane boundary (see App.~\ref{notation} for details). Using the transformation laws \eqref{sym1} one can easily check that the purely tangential components to the worldvolume of $B^{\mu\nu\rho}$ are in fact a gauge artifact,
\beq
\delta_{1}(B^{\mu\nu\rho}u_{\rho}^{a})=\epsilon^{\mu\nu a}~.
\eeq
Hence, the components $B^{\mu\nu a}$ can be gauged away everywhere except at the boundary where the parameters $\epsilon^{\mu\nu a}$ cannot be freely chosen. This implies that there are degrees of freedom that live exclusively on the boundary of the worldvolume, for which a physical interpretation will be given in the next section.

\subsubsection*{Extra symmetry 2}
The stress-energy tensor $T^{\mu\nu}$ has been expanded around the surface $x^{\mu}=X^{\mu}(\sigma^{a})$ as in \eqref{stpd} but since we are dealing with objects of finite thickness there is freedom in choosing a different worldvolume. In physical terms, the finite thickness of the brane allows for different choices of worldvolume surfaces\footnote{In the particle case there is a natural choice of reference frame which is the centre of mass.}. This redundancy is an exact symmetry of the full series expansion \eqref{stpd} but only an approximate one of the truncated series to order $k=1$. This is because neglecting higher order terms in the expansion \eqref{stpd} already constrains the allowed choices of worldvolumes. In particular, choosing the surface $X^{\alpha}(\sigma^a)$ to lie outside the localized matter would require a non-zero contribution from the higher order $B$-tensors. Therefore, we choose the surface to lie within the localized matter and assume the following hierarchy of scales:
\beq
B^{\mu\nu}=\mathcal{O}_0,\quad\quad B^{\mu\nu\rho}=\mathcal{O}_1,\quad\quad B^{\mu\nu\rho\lambda}=\mathcal{O}_2,\quad ...~.
\eeq
In this way, we can define the action of `extra symmetry 2' as
\beq \label{sym2}
X^{'\alpha}(\sigma^a)=X^{\alpha}(\sigma^a)+\epsilon^{\alpha}(\sigma^a)~,
\eeq
where $\epsilon^{\alpha}$ is constrained by the requirement that the transformed $B$-tensors obey $B_{k+1}=\mathcal{O}_{k+1}$. In both the single-pole and pole-dipole cases this implies $\epsilon^{\alpha}=\mathcal{O}_1$. The action of \eqref{sym2} to order $k=1$ demands the following transformation rule for the $B$-tensors:
\beq
\delta_2 B^{\mu\nu}=-B^{\mu\nu}u^{a}_{\rho}\nabla_{a}\epsilon^{\rho}-2B^{\lambda(\mu}\Gamma^{\nu)}_{\lambda\rho}\epsilon^{\rho},\quad\quad \delta_2 B^{\mu\nu\rho}=-B^{\mu\nu}\epsilon^{\rho}~,
\label{symm2}\eeq
where we have ignored contributions of $\mathcal{O}_2$ and higher. In the single-pole approximation we find $\delta_2 X^{\alpha}=0$ and $\delta_2 B^{\mu\nu}=0$, emphasizing the fact that there is no freedom in choosing the worldvolume surface for the object as they are infinitely thin.

\subsection{Equations of motion and physical interpretation}
The equations of motion (EOMs) for probe pole-dipole branes moving in a curved background spacetime can be obtained by solving Eq.~\eqref{emc} using the stress-energy tensor given in \eqref{stpd} truncated to order $k=1$. The derivation of these equations is somewhat involved and we refer to \cite{Vasilic:2007wp} for an extensive detailed analysis. 

It is convenient to decompose the objects $B^{\mu\nu}$ and $B^{\mu\nu\rho}$ into tangential and orthogonal components to the worldvolume, the latter being subjected to the constraint equation \cite{Vasilic:2007wp}
\beq
{\perp^{\nu}}_{\lambda}{\perp^{\sigma}}_{\rho}B^{\mu(\lambda\rho)}=0~,
\eeq
while the former is not altogether independent and bares a relation with $B^{\mu\nu\rho}$ which will be described below. This suggests the following decomposition\footnote{A subindex $_\perp$ on a tensor indicates that all $\mu, \nu$ type of indices are orthogonal, e.g., $B_\perp^{a\mu}={\perp^\mu}_\nu B_\perp^{a\nu}$. Details can be found in App.~\ref{notation}.}:
\beq \label{bb}
\begin{split}
B^{\mu\nu}=B^{\mu\nu}_{\perp}+2u_{b}^{(\mu}B^{\nu)b}_{\perp}+u^{\mu}_a u^{\nu}_b B^{ab},\quad\quad B^{\mu\nu\rho}=2u_{b}^{(\mu}B^{\nu)\rho b}_{\perp}+u^{\mu}_a u^{\nu}_b B^{\rho ab}_{\perp}+u^{\rho}_{a}B^{\mu\nu a}~,
\end{split}
\eeq
that can be shown to obey the properties $B_{\perp}^{(\mu\nu)a}=B^{\mu[ab]}_{\perp}=B^{[\mu\nu]a}=0$. Due to `extra symmetry 1', described in the previous section, the last components in the decomposition of $B^{\mu\nu\rho}$ are left neither parallel nor perpendicular to the worldvolume, as $B^{\mu\nu a}$ can be gauged away in the bulk of the brane. A convenient form of the EOMs can be obtained by defining a new set of tensors
\beq \label{sn}
S^{\mu\nu a}=B^{\mu\nu a}_{\perp}+u_{b}^{[\mu}B^{\nu]ba}_{\perp},\quad\quad N^{\mu\nu a}=B^{\mu\nu a}+u_{b}^{(\mu}B^{\nu)ba}_{\perp}~,
\eeq
which are, respectively, anti-symmetric and symmetric in the first two indices $\mu,\nu$. In terms of these it is straightforward to check that $B^{\mu\nu\rho}$ can be recast as
\beq \label{bbsn}
B^{\mu\nu\rho}=2u_{a}^{(u}S^{\nu)\rho a}+N^{\mu\nu a}u_{a}^{\rho}~.
\eeq
Furthermore, the interdependence between the orthogonal components of $B^{\mu\nu}$ and the quantities $S^{\mu\nu a}$ and $N^{\mu\nu a}$ is expressed through the relations
\beq \label{bsn}
B^{\mu\nu}_{\perp}={\perp^{\mu}}_{\lambda}{\perp^{\nu}}_{\rho}\nabla_{a}N^{\lambda \rho a},\quad \quad B^{\mu a}_{\perp}=u^{a}_{\lambda}{\perp^{\mu}}_{\rho}\nabla_{b}\left(S^{\lambda\rho b}+N^{\lambda\rho b}\right)~,
\eeq
while the tangential components $B^{ab}$ describe the monopole contribution to the intrinsic stress-energy tensor of the brane.

Parametrizing the EOMs using \eqref{bbsn}-\eqref{bsn} yields two sets of bulk equations: a partial conservation equation of the brane worldvolume currents $S^{\mu\nu a}$,
\beq \label{sc}
{\perp^{\mu}}_{\lambda}{\perp^{\nu}}_{\rho}\nabla_{a}S^{\lambda\rho a}=0~,
\eeq
and the equation that describes the motion of the pole-dipole brane
\beq\label{wv1}
\nabla_{b}\left(m^{ab}u_{a}^{\mu}-2u^{b}_{\lambda}\nabla_{a}S^{\mu\lambda a}+u^{\mu}_{c}u^{c}_{\rho}u^{b}_{\lambda}\nabla_{a}S^{\rho\lambda a}\right)-u_{a}^{\nu}S^{\lambda\rho a}{R^{\mu}}_{\nu\lambda\rho}&=0~,
\eeq
where we have defined, for later convenience, the worldvolume tensor $m^{ab}$ through the formula
\beq
m^{ab}=B^{ab}-u^{a}_{\rho}u^{b}_{\lambda}\nabla_{c}N^{\rho\lambda c}~.
\eeq
Eqs.~\eqref{sc}-\eqref{wv1} reduce to those of a spinning point particle as obtained by Papapetrou in \cite{Papapetrou:1951pa} when $p=0$. In order to highlight the physical meaning of Eq.~\eqref{wv1}, we project it along the tangential and orthogonal directions to the worldvolume. This operation leads to the intrinsic and extrinsic worldvolume equations:
\beq\label{int1}
\nabla_{b}m^{ab}=2\nabla_{b}\left(u_{\rho}^{[b}{K^{a]}}_{c\lambda}S^{\rho\lambda c}\right)-2u_{\lambda}^{b}{K^{a}}_{b\rho}\nabla_{c}S^{\rho\lambda c}-u_{\mu}^{a}u^{\nu}_{c}S^{\rho\lambda c}{R^{\mu}}_{\nu\lambda\rho}~,
\eeq
\beq\label{ext1}
m^{ab}{K_{ab}}^{\rho}=\left(2{\perp^{\rho}}_{\lambda}{K^{b}}_{b\nu}+u_{\nu}^{c}u_{\lambda}^{b}{K_{bc}}^{\rho}\right)\nabla_{a}S^{\nu\lambda a}+ 2u_{\nu}^{b}{\perp^{\rho}}_{\lambda}\nabla_{b}\nabla_{a}S^{\nu\lambda a}- u_{a}^{\nu}{\perp^{\rho}}_{_{ \mu}}S^{\sigma\lambda a}{R^{\mu}}_{\nu\lambda\sigma}~.
\eeq
In this way, it is clear that Eq.~\eqref{int1} can be interpreted as an equation for the conservation of the intrinsic monopole stress-energy tensor $B^{ab}$, which can be violated due to the higher order dipole contributions\footnote{Even though the conservation of the monopole stress-energy tensor is not necessarily guaranteed, in the cases studied here we always find that $B^{ab}$ is conserved.}, while Eq.~\eqref{ext1} is the generalized geodesic equation for a pole-dipole $p$-dimensional object, in contrast with the single-pole case \eqref{bf1}. 

In turn, the EOMs that govern the brane dynamics \eqref{sc}-\eqref{wv1} everywhere inside $\mathcal{W}_{p+1}$ are supplemented by well defined boundary conditions derived from solving Eq.~\eqref{emc},
\beq \label{bound}
\begin{split}
S^{\mu\nu a}\hat{n}_{a}\hat{n}_{\nu}|_{\partial\mathcal{W}_{p+1}}&=0\\
{\perp^{\mu}}_{\lambda}{\perp^{\nu}}_{\rho}S^{\lambda\rho a}\hat{n}_{a}|_{\partial\mathcal{W}_{p+1}}&=0\\
\left[\nabla_{\hat{i}}\left(N^{\hat{i}\hat{j}}v_{\hat{j}}^{\mu}+2S^{\mu\nu a}\hat{n}_{a}v_{\nu}^{\hat{i}}\right)-\hat{n}_{b}\left(m^{ab}u_{a}^{\mu}-2u^{b}_{\lambda}\nabla_{a}S^{\mu\lambda a}+u^{\mu}_{c}u^{c}_{\rho}u^{b}_{\lambda}\nabla_{a}S^{\rho\lambda a}\right)\right]|_{\partial\mathcal{W}_{p+1}}&=0~,
\end{split}
\eeq
where $v_{\nu}^{\hat{j}}$ are the boundary coordinate vectors (see App.~\ref{notation} for details) and we have defined $N^{\hat{i}\hat{j}}=N^{\mu\nu a}\hat{n}_a v_{\mu}^{\hat{i}}v_{\nu}^{\hat{j}}$. These quantities appear only in the boundary conditions and nowhere else and, as mentioned in the previous section while discussing `extra symmetry 1', contain the degrees of freedom that live exclusively on the boundary. In full generality, the tensors $m^{ab}$, $S^{\mu\nu a}$ and $N^{\hat{i}\hat{j}}$ characterize the internal structure of the brane and play a crucial role in describing its dynamics. We will proceed by analyzing their physical meaning and of the resulting EOMs.

\subsubsection*{Physical interpretation}
Ref.~\cite{Vasilic:2007wp} introduced the $S^{\mu\nu a}$ and $m^{ab}$ quantities, in terms of which the equations of motion of pole-dipole branes simplify considerably. In the following we give an interpretation of $S^{\mu\nu a}$, and see that it contains two types of contributions: the genuine intrinsic transverse angular momenta, and the dipole moment of the distribution of worldvolume stress-energy\footnote{In the case of the $0-$brane the dipole can be gauged away and, as demonstrated in the original work of Corinaldesi and Papapetrou \cite{Papapetrou:1951pa, Corinaldesi:1951pb}, $S^{\mu\nu a}$ describes only spin degrees of freedom.}. Thus, besides generalising Papapetrou's equations, Eqs.~\eqref{int1}, \eqref{int2} include dipole interactions analogous to those in \eqref{poledipoleeom}.

We begin by analyzing which components of $S^{\mu\nu a}$ are involved in the description of the intrinsic angular momenta. To this end, we assume to be working in flat spacetime written in Cartesian coordinates and focus on uniform $p$-branes extended along the $x^{0},...,x^{p}$ directions. Evaluating the total angular momentum on the transverse plane labeled by the indices $\mu,\nu$ leads to
\beq \label{jflat}
J^{\mu\nu}_{\perp}=\int_{\Sigma} d^{D-1}x \left(T^{0\mu}x^{\nu}-T^{0\nu}x^\mu\right)=\int_{\mathcal{B}_{p}} d^p \sigma\sqrt{-\gamma} \left(2B^{0\mu\nu}_{\perp}\right)+\text{boundary terms}~,
\eeq
where $\Sigma$ is a constant time slice in the bulk spacetime. At this point, we ignore the boundary terms, which only depend on the components $B^{\mu\nu a}$, but we will consider them towards the end of this section. From \eqref{jflat}, we can see that the monopole contribution to the intrinsic stress-energy tensor $B^{ab}$ does not play a role in \eqref{jflat} and hence a $p$-brane when treated in the single-pole approximation can never carry intrinsic angular momenta. Furthermore, only the components $B^{a\mu\nu}_{\perp}$ contain information about the spin of the object. This suggests the introduction of a current density of transverse angular momenta as
\beq \label{j}
j^{a\mu\nu}=2u^a_\rho {\perp^\mu}_\sigma{\perp^\nu}_\lambda B^{\rho[\sigma\lambda]}=2B^{a\mu\nu}_{\perp}~,
\eeq
where both indices $\mu,\nu$ are orthogonal to the worldvolume. 

On the other hand, there is another source of $B^{\mu\nu\rho}$ which is of a different nature than transverse angular momenta. It arises from the fact that, since we are probing the finite thickness of the brane, we need to take into account corrections to the intrinsic stress-energy tensor $T^{ab}$ due to the dipole-type effects. This is characterized by the integral\footnote{As a matter of a fact, this is the usual notion of an electric induced dipole. In electrostatics, given a density of charge $\rho(x)$, the dipole can computed as,
\beq \nonumber
\vec{D}=\int_\Sigma d^{D-1}x \vec{x}\rho(x)~.
\eeq},
\beq\label{Dflat}
D^{ab\rho}=\int_\Sigma d^{D-1}x T^{ab}x^\rho=\int_\Sigma d^{D-1}x T^{\mu\nu}u^{a}_{\mu}u^{b}_{\nu}x^\rho= \int_{B_p} d^p\sigma \sqrt{-\gamma}B_{\perp}^{\rho ab}+\text{boundary terms}~,
\eeq
where $x^{\rho}$ is an orthogonal coordinate to the worldvolume. $D^{ab\rho}$ captures the dipole moment of the distribution of worldvolume stress-energy. As in the case of intrinsic angular momenta, we introduce a current density that describes such deformations to the intrinsic stress-energy tensor by
\beq \label{d}
d^{ab\rho}=u_{\mu}^{a}u_{\nu}^{b}{\perp^{\rho}}_{\lambda}B^{\mu\nu\lambda}=B_{\perp}^{\rho ab}~,
\eeq
where the index $\rho$ is orthogonal to the worldvolume $\mathcal{W}_{p+1}$.

Our aim now is to recast the EOMs, including the boundary conditions, in terms of these newly defined quantities. Using the definitions of the current densities \eqref{j} and \eqref{d}, we can rewrite the tensors introduced in \eqref{sn} as
\beq \label{nsn}
S^{\mu\nu a}=\frac{1}{2}j^{a\mu\nu}-d^{ab[\mu} u_b^{\nu]},\quad\quad N^{\mu\nu a}=B^{\mu\nu a}+d^{ab(\mu} u_b^{\nu)}~.
\eeq
We note that we have not been concerned so far with giving a physical interpretation to the components $B^{\mu\nu a}$. This is because, due to `extra symmetry 1', we can gauge them away everywhere in the bulk while on the boundary we will have to deal with $N^{\hat{i}\hat{j}}$ as we will see below. In turn, the current conservation equation \eqref{sc} becomes:
\beq \label{jc}
\frac{1}{2}{\perp^\mu}_\lambda{\perp^\nu}_\rho\nabla_a j^{a\rho\lambda}+d^{ab\left[\nu\right.}{K_{ab}}^{\left.\mu\right]}=0~.
\eeq
This equation can be interpreted as the balance between orbital angular momentum and intrinsic angular momentum. Nevertheless, as it will be argued in Sec.~\ref{YMandRings}, for blackfold-type objects, the dipole current $d^{ab \rho}$ is induced by the extrinsic curvature. In all such situations, the second term in Eq.~\eqref{jc} vanishes, leading to a conserved spin current which can be naturally interpreted as a 0-brane particle current on the worldvolume.

We now turn our attention to the intrinsic and extrinsic Eqs.~\eqref{int1}-\eqref{ext1}, which can be rewritten using \eqref{j} and \eqref{d} as
\beq\label{int2}
D_a m^{ab}={{K_c}^b}_\mu\left({{K_a}^c}_\lambda j^{a\mu\lambda}+\nabla_a d^{ac\mu}\right)+\nabla_c{{K_a}^{\left[b\right.}}_\mu d^{\left.c\right]a\mu}-u_\mu^bu^\nu_a \left(\frac{1}{2}j^{a\lambda\rho}-d^{ab[\lambda} u_b^{\rho]}  \right){R^\mu}_{\nu\lambda\rho}~,
\eeq
\begin{eqnarray}
 \label{ext2}
m^{ab}{K_{ab}}^\rho&=&\nabla_{b}\left(j^{a\lambda\rho}{K^{b}}_{a\lambda}\right)+u^{\rho}_{c}{K^{a}}_{b\lambda}{K^{c}}_{a\sigma}j^{b\lambda\sigma}-{K_{ac}}^{\rho}{K^{(c}}_{b\lambda}d^{a)b\lambda}-{\perp^{\rho}}_{\sigma}\nabla_{b}\nabla_{a}d^{ab\sigma} \nn \\ && - ~ u_{a}^{\nu}{\perp^{\rho}}_{\mu}\left(\frac{1}{2}j^{a\lambda\sigma}-d^{ab[\lambda} u_b^{\sigma]}  \right){R^{\mu}}_{\nu\lambda\sigma}~.
\end{eqnarray}
These equations provide the gravitational analog of Eq.~\eqref{poledipoleeom} for $p$-branes. Written in this way it is apparent that, besides spin interactions, we also have couplings to the dipole current $d^{ab\rho}$. For blackfold objects these interactions can be interpreted as elastic forces, for which a derivation in terms of high-pressure elasticity theory \cite{Quintana:1972} can in principle be accomplished and will be presented elsewhere. This point will be further motivated in Sec.~\ref{YMandRings}. Finally, the boundary conditions \eqref{bound} take the form:
\beq
\begin{split} \label{bound2}
d^{ab\mu}\hat{n}_{a}\hat{n}_{b}|_{\partial\mathcal{W}_{p+1}}&=0 \\
j^{a\mu\nu}\hat{n}_{a}|_{\partial \mathcal{W}_{p+1}}&=0\\
\left[\nabla_{\hat{i}}\left(N^{\hat{i}\hat{j}}v_{\hat{j}}^{\mu}-d^{ab\mu}\hat{n}_{a}v^{\hat{i}}_{b}\right)-\hat{n}_b\left(m^{ab}u_{a}^{\mu}-j^{a\mu\lambda}{{K_a}^b}_\lambda-{{K_a}^{\left(c\right.}}_\lambda d^{\left.b\right)a\lambda}u_c^{\mu}-\nabla_a d^{ab\mu}\right)\right]|_{\partial \mathcal{W}_{p+1}}&=0~,
\end{split}
\eeq
where $N^{\hat{i}\hat{j}}v_{\hat{j}}^{\mu}$ is in fact
\beq
N^{\hat{i}\hat{j}}v_{\hat{j}}^{\mu}=B^{\lambda\nu a}\hat{n}_{a}v^{\hat{i}}_{\lambda}v^{\hat{j}}_{\nu}v^{\mu}_{\hat{j}}~.
\eeq
Eqs.~\eqref{jc}-\eqref{bound2} when applied to black $p$-branes constitute the blackfold equations in the pole-dipole approximation and will be analyzed in detail in particular cases throughout the course of this work.

We end this section by briefly commenting on the physical interpretation of the coefficients $N^{\hat{i}\hat{j}}$. As remarked in \cite{Vasilic:2007wp}, $N^{\hat{i}\hat{j}}$ characterizes the tangential components to the worldvolume of the brane thickness. The reason why these components drop out of the bulk equations is because thickening the brane along tangential directions does not affect the brane interior but if the brane has a boundary then it will be affected by such process. Essentially, $N^{\hat{i}\hat{j}}$ is a correction to the intrinsic stress-energy tensor of the brane boundary $T^{\hat{i}\hat{j}}$ and can be read off from an analytic solution by evaluating the stress-energy tensor on the boundary of the brane surface. Since for all the cases analyzed in this paper $B^{\mu\nu a}$ vanishes everywhere, including at the boundary, we assume $B^{\mu\nu a}=0$ from hereon.

\subsection{Physical properties} \label{thermo}
Branes in the pole-dipole approximation can carry different conserved charges and this section is devoted to describing them. In the case of blackfolds, where these charges acquire a thermodynamic interpretation, we present a method for obtaining the remaining thermodynamic quantities involved. 

As mentioned in Sec.~\ref{intro}, fine structure corrections do not violate stress-energy conservation \eqref{emc}. Hence, since we are working in the probe approximation, associated with any background Killing vector field $\textbf{k}^{\mu}$, there exists a conserved current given by $T^{\mu\nu}\textbf{k}_{\nu}$ and thus the conserved charges can be obtained in the usual way. In general, we have a charge $Q_\textbf{k}$ given by
\beq \label{qc}
|Q_{\textbf{k}}|=\int_{\mathcal{B}_{p}}dV_{(p)}B^{\mu\nu}n_{\mu}\textbf{k}_{\nu}+\int_{\mathcal{B}_p}dV_{(p)}B^{\mu\nu\rho}\nabla_{\rho}\left(n_{\mu}\textbf{k}_{\nu}\right)~,
\eeq
where $n^{\mu}$ is the normal vector to a constant time slice of the background spacetime and is defined as
\beq
n^{\mu}=\frac{\xi^{\nu}}{R_{0}},~~~~R_{0}=\sqrt{-\xi^2}|_{\mathcal{W}_{p+1}}~.
\eeq
Here, $\xi^{\nu}$ is the Killing vector field associated with asymptotic time translations and $R_{0}$ is the redshift factor. In writing Eq.~\eqref{qc} we have assumed that $\xi^{\nu}$ is hypersurface orthogonal with respect to the background spacetime and that $\xi^{\nu}$ is parallel to the worldvolume timelike Killing vector field, which is also hypersurface orthogonal with respect to the worldvolume metric, i.e., $\xi^{\mu}=u^{\mu}_{a}\xi^{a}$. 

Using the decompositions \eqref{bb} and Eqs.~\eqref{j},\eqref{d} we can write down general expressions for the physical quantities in terms of the spin and dipole currents. The total mass, associated with $\xi^{\nu}$, reads\footnote{Here we have used the Killing equation $\nabla_{(\mu}\textbf{k}_{\mu)}=0$ to exchange covariant derivatives for partial derivatives.}
\beq \label{m}
M=\int_{\mathcal{B}_p}dV_{(p)}B^{ab}n_{a}\xi_{b}+\int_{\mathcal{B}_p}dV_{(p)}d^{ab\rho}\xi_{a}u^{\mu}_{b}\partial_{\rho}n_\mu~.
\eeq
The first term above describes the contribution to the total mass arising from a monopole source of stress-energy, while the second is the extra contribution coming from a dipole distribution. Similarly, the total angular momentum along a worldvolume spatial direction $i$ associated with the rotational Killing vector field $\chi^{i}$ takes the form
\beq\label{j1}
J^{i}=-\int_{\mathcal{B}_p}dV_{(p)}B^{ab}n_{a}\chi_{b}^{i}-\frac{1}{2}\int_{\mathcal{B}_p}dV_{(p)}d^{ab\rho}u_{b}^{\mu}\left(\xi_{a}\partial_{\rho}\frac{\chi^{i}_{\mu}}{R_0}+\chi^{i}_{a}\partial_{\rho}n_{\mu}\right)~.
\eeq
Moreover, the transverse angular momentum associated with the rotational Killing vector field $\chi^{\alpha}_{\perp}$ is given by
\beq \label{jt}
J^{\alpha}_{\perp}=-\int_{\mathcal{B}_p}dV_{( p)}\left(\frac{1}{2}\nabla_{a}j^{a\lambda\rho}+u_{b}^{\lambda}\nabla_{a}d^{ab\rho}\right)n_{\lambda}\chi_{\perp\rho}-\frac{1}{4}\int_{\mathcal{B}_p}dV_{(p)}j^{a\mu\rho}u_{a}^{\nu}\left(\xi_{\nu}\partial_{\rho}\frac{\chi^{\alpha}_{\perp\mu}}{R_0}+\chi_{\perp \mu}^{\alpha}\partial_{\rho}n_{\nu}\right)\,.
\eeq
We see that the spin current $j^{a\mu\rho}$ only plays a role in the transverse angular momenta. On the other hand, the dipole current $d^{ab\rho}$ does influence all charges, including $J^{\alpha}_{\perp}$ (which corresponds to orbital angular momentum). This is because, in general, $j^{a\mu\rho}$ is not a conserved current due to Eq.~\eqref{jc}. In the cases considered in this paper, the first term in Eq.~\eqref{jt} always vanishes, as the spin current is always conserved.

\subsubsection*{Thermodynamic quantities}
When considering the dynamics of black $p$-branes, the conserved charges have a thermodynamic interpretation. Furthermore, a local temperature and entropy density can be assigned to the $p$-dimensional object from which one can then compute a global temperature and total area. As we will be mainly concerned with blackfolds constructed from wrapped Schwarzschild and Myers-Perry $p$-branes on curved submanifolds, the knowledge of these quantities amounts to perform a near-horizon computation which is beyond the scope of this paper. Instead, we present here an alternative method to obtain the on-shell temperature and entropy. 

Focusing for the moment on asymptotically flat spacetimes, the method consists of using the first law of black hole thermodynamics
\beq\label{1law}
dM=\sum_{i}\Omega_{i}dJ^{i}+\sum_{\alpha}\Omega_{\alpha}dJ_{\perp}^{\alpha} + TdS~,
\eeq
together with the Smarr relation
\beq \label{smarr}
(n+p)M=(n+p+1)\left(\sum_{i}\Omega_{i}J^{i}+\sum_{\alpha}\Omega_{\alpha}J_{\perp}^{\alpha}+TS\right)~,
\eeq
in order to determine the two unknown quantities $T$ and $S$. Since, in general, all physical quantities depend on a set of intrinsic parameters, such as $r_0$ and $R$, which we collectively call by $\Phi^{w}$, then, by means of Eq.~\eqref{smarr} the product $TS$ can be determined, as all other quantities involved can be evaluated using the expressions \eqref{m}-\eqref{jt}. Inserting this result into Eq.~\eqref{1law} leads to a set of equations, one for each $\Phi^{w}$, which take the following form:
\beq \label{thermosets}
\frac{1}{T}\frac{\partial T}{\partial \Phi^{w}}=\frac{1}{TS}\left(\frac{\partial TS}{\partial \Phi^{w}}+\sum_{i}\Omega_{i}\frac{\partial J^{i}}{\partial\Phi^{w}}+\sum_{\alpha}\Omega_{\alpha}\frac{\partial J^{\alpha}_{\perp}}{\partial \Phi^{w}}-\frac{\partial M}{\partial \Phi^{w}}\right)~.
\eeq
Solving this set of equations yields the temperature of the black object and hence also the entropy by Eq.~\eqref{smarr}. The method just described provides the knowledge of $T$ and $S$ up to a constant, which can later be fixed by demanding the correct behavior in the infinitely thin limit, or in the single-pole approximation, where all quantities can be unambiguously determined using the results of Ref.~\cite{Emparan:2009at}. 

We conclude this section by briefly considering blackfolds in (A)dS backgrounds in situations where the dipole current $d^{ab\rho}$ vanishes. In such cases the total integrated tension $\mathcal{T}$, given by
\beq \label{tension}
\mathcal{T}=-\int_{\mathcal{B}_{p}} dV_{(p)} R_{0}B^{\mu\nu}\left(h_{\mu\nu}-n_{\mu}n_{\nu}\right)-\int_{\mathcal{B}_{p}} dV_{(p)}R_{0} B^{\mu\nu\rho}\nabla_{\rho}\left(h_{\mu\nu}-n_{\mu}n_{\nu}\right)~,
\eeq
plays a role in the thermodynamics and must be added to the rhs of Eq.~\eqref{smarr}. These considerations will be used later in Sec.~\eqref{dsbh} and App.~\eqref{kerrads} in order to make contact with the thermodynamic properties of doubly-spinning Myers-Perry and higher-dimensional Kerr-(A)dS black holes.


\section{Black string elasticity and corrections to thin black rings}\label{YMandRings}
The purpose of this section is to study how a Schwarzschild black string reacts to its bending. This is done using an explicit solution describing a black string of thickness $r_0$ bent on a circle of radius $R$, to first order in $r_0/R$. Very similarly to what happens to elastic rods, the effective blackfold stress-energy tensor distribution acquires a dipole contribution, exhibiting an effective elastic behavior.
\subsection{Measuring the dipole from the approximate analytic solution}\label{CalculationDipole}
This section uses the solutions constructed in \cite{Emparan:2007wm} to compute the dipole contribution to the effective stress-energy distribution induced by the bending of strings. This reference computes, to first order in $r_0/R$, the spacetime of a bent Schwarzschild string, where $r_0$ is the thickness of the string and $R$ is the radius of curvature of the circle on which the string is bent. In Ref.~\cite{Emparan:2007wm} this geometry is obtained by solving Einstein's equations using a matched asymptotic expansion (MAE), in which there are two different coordinate patches. These two zones are the zone near the horizon $r\ll R$, and the far zone $r\gg r_0$, where the weak field approximation applies. The applicability of a MAE requires that there is an overlap zone, and this happens when $r_0\ll R$, which is why this technique is amenable in this regime of parameters.

In the near zone, the geometry is a perturbation of the Schwarzschild black string, and in the far region the gravitational field is well described by the linear approximation sourced by the appropriate blackfold effective probe distribution of stress-energy. In MAE each zone feeds the other zone with boundary conditions.

Let us summarize the order-by-order logic at each zone. We denote by {\bf $k$th (near/far)} the geometry to order $(r_0/R)^k$ in each of the zones. The first few orders go as follows:
\begin{itemize}
\item {\bf 0th (near/far):} The geometry is that of the straight boosted Schwarzschild black string: 
\beq\begin{split}
ds^2=&\frac{dr^2}{1-\frac{r_0^n}{r^n}}+r^2\left(d\theta^2+\sin^2\theta d\Omega_{(n)}^2\right)\\
&-\left(1-\frac{r_0^n}{r^n}\right)\left(\cosh\alpha\, dt +\sinh\alpha\, dz\right)^2+\left(\cosh\alpha\, dz+\sinh\alpha\, dt\right)^2\,,
\end{split}
\eeq
and the far field is well described by the linearized approximation, $\Box \bar{h}_{\mu\nu}=-16\pi G T_{\mu\nu}$, sourced by the ADM stress-energy tensor of the string, which is localized on its effective worldvolume \eqref{monopoletensor}, $T_{\mu\nu}=B_{\mu\nu}\delta^{n+2}(r)$:
\begin{subequations}
\beq
B_{tt}=\frac{\Omega_{(n+1)}r_0^n}{16\pi G}\left(n\cosh^2\alpha+1\right)\,,
\eeq
\beq
B_{tz}=\frac{\Omega_{(n+1)}r_0^n}{16\pi G}n\cosh\alpha\sinh\alpha\,,
\eeq
\beq
B_{zz}=\frac{\Omega_{(n+1)}r_0^n}{16\pi G}\left(n\sinh^2\alpha-1\right)\,.
\eeq
\label{Tmunu}\end{subequations}
\item {\bf 1st (far):} The far field is sourced by the blackfold effective stress-energy tensor \eqref{Tmunu}, now along a bent manifold with curvature $1/R$. Only sources satisfying  $\nabla^\mu T_{\mu\nu}=0$ can be coupled to the gravitational field, and this conservation equation constitutes the blackfold effective equations of motion. For the case of the ring they reduce to constancy of $r_0$ and $B_{zz}=0$, which sets the rapidity to
\beq
\sinh^2\alpha=\frac{1}{n}\,.
\label{leadingequilrings}\eeq
This is interpreted as the value at which the centrifugal repulsion compensates the tension.
\item {\bf 1st (near):} The geometry in the near zone is a $1/R$ perturbation to the Schwarzschild black string, that is found using {\bf 1st (far)} as a boundary condition. One can compute, from the far field of the corrected near zone solution, a corrected effective stress-energy source in the far zone.
\item {\bf 2nd (far):} One would use the corrected effective distribution of stress-energy measured in {\bf 1st (near)} to source the far field. The corrected stress-energy tensor would need to  satisfy a corrected effective equation of motion coming from $\nabla_\mu T^{\mu\nu}$, which would modify \eqref{leadingequilrings}.
\item {\bf 2nd (near):} One would use the result of {\bf 2nd (far)}, as a boundary condition to solve the near geometry as a $1/R^2$ perturbation of the black string. From this near solution, one would extract yet another correction to the source of the far field which would modify the effective equation of motion...
\end{itemize}
and so on. \cite{Emparan:2007wm} went to the step {\bf 1st (near)}. In this section we push that analysis to {\bf 2nd (far)} by reading the correction to the stress-energy source, which is of dipole type, and which will have the equations derived in Sec.~\ref{bf2} as effective equation of motion.

The solution of a bent black string in flat space at {\bf 1st (near)} was written down in \cite{Emparan:2007wm} as
\begin{subequations}\label{gmunu}
\begin{equation}
\label{gtta}
g_{tt} = - 1 + \frac{n+1}{n} \frac{r_0^{n}}{r^{n}} +
\frac{\cos \theta}{R} a(r)\,,
\end{equation}
\begin{equation}
g_{tz} = \frac{\sqrt{n+1}}{n} \left[ \frac{r_0^{n}}{r^{n}} +
\frac{\cos \theta}{R} b(r) \right]\,,
\end{equation}
\begin{equation}
g_{zz} = 1 + \frac{1}{n} \frac{r_0^{n}}{r^{n}} + \frac{\cos \theta}{R}
c(r)\,,
\end{equation}
\begin{equation}\label{grr}
g_{rr} = \left( 1- \frac{r_0^{n}}{r^{n}} \right)^{-1} \left[ 1
+ \frac{\cos \theta}{R} f(r) \right]\,,
\end{equation}
\begin{equation}
\label{gijg}
g_{ij} =  \hat g_{ij}\left[ 1
+ \frac{\cos \theta}{R} g(r) \right]\,,
\end{equation}
\end{subequations}
where $\hat{g}_{ij}$ is the metric of a round $n+1$ sphere of radius $r$ in polar coordinates:
\beq
\hat{g}_{ij} dx^i dx^j=r^2 d\Omega_{(n+1)}^2=r^2\left(d\theta^2+\sin^2\theta\, d\Omega^2_{(n)}\right)\,.
\eeq
As expected, the limit $R\rightarrow\infty$ corresponds to a black string boosted in the $z$ direction with rapidity $\sinh\alpha=1/\sqrt{n}$, according to \eqref{leadingequilrings}.

The regular solution to the Einstein equations  found in \cite{Emparan:2007wm} has the following large $r$ asymptotics
\begin{subequations}\label{aasy}
\beq
a(r)=\left[k_1(1+n)-n(4+3n)\xi(n)\right]\frac{r_0^{n+2}}{r^{n+1}}+\mathcal{O}\left(r^{-(n+2)}\right)\,,
\eeq
\beq
b(r)=\frac{r_0^n}{r^{n-1}}+\left[k_1 n-2n^2\xi(n)\right]\frac{r_0^{n+2}}{r^{n+1}}+\mathcal{O}\left(r^{-(n+2)}\right)\,,
\eeq
\beq
c(r)=2r+\frac{1}{n}\frac{r_0^n}{r^{n-1}}+k_1\frac{r_0^{n+2}}{r^{n+1}}- \frac{n^2+2n-1}{2n^2(n-2)}\frac{r_0^{2n}}{r^{2n-1}}+k_2\frac{r_0^{2n+2}}{r^{2n+1}}+\mathcal{O}\left(r^{-(2n+2)}\right)
\,,
\label{Expansionc}\eeq
\beq
f(r)=-\frac{2}{n}r+\frac{1}{n^2}\frac{r_0^n}{r^{n-1}}+\left[(k_1+2k_2)n-(4n^2+7n+4)\xi(n)\right]\frac{r_0^{n+2}}{r^{n+1}}+\mathcal{O}\left(r^{-(n+2)}\right)\,,
\eeq
\beq
g(r)=-\frac{2}{n}r+\frac{1}{n^2}\frac{r_0^n}{r^{n-1}}+\left[\frac{2(k_1-k_2 n)}{1+n}+(n-4)\xi(n)\right]\frac{r_0^{n+2}}{r^{n+1}}+\mathcal{O}\left(r^{-(n+2)}\right)\,,
\eeq
\end{subequations}
where we defined
\beq
\xi(n)=-\frac{2^{-\frac{4+n}{n}}}{n+1}\frac{\Gamma\left(\frac{2n+1}{n}\right)\Gamma\left(-\frac{n+2}{2n}\right)}{\Gamma\left(-\frac{n+1}{n}\right)\Gamma\left(\frac{n+2}{2n}\right)}\,,
\eeq
which is zero for $n=1$ and divergent for $n=2$. In the latter case this is because the corresponding terms in the expansion should have logarithm contributions on top the polynomial ones. The interest of this paper is in the $n>2$ cases, which are those in which the step {\bf 2nd (far)} can be interpreted in the probe approximation, as discussed in the introduction. In what follows we assume $n>2$, \ie black rings in more than six dimensions.

It is worth mentioning various things about expressions \eqref{aasy}. First, note that all functions are expanded to order $r^{-(n+2)}$ except for $c(r)$, which is expanded to order $r^{-(2n+2)}$. 

Second, there is residual gauge freedom in the coordinate system in which the metric reads \eqref{gmunu} and, because of that, $c(r)$ is essentially unfixed by the Einstein equations. This shows up in the unfixed $k_1$ and $k_2$ parameters, which parametrize gauge ambiguities. $c(r)$ is not completely free because we demand that at large $r$ the metric goes to a known metric in a familiar coordinate system. That known metric is the one that one recovers if one keeps only terms in this expansions up to order $r^{-n}$ in \eqref{gmunu}. The geometry to order $r^{-n}$ is just the linearized gravitational field of the blackfold distributional stress-energy tensor sitting on a circle of radius $R$, to first order in $1/R$.

To be more specific with this interpretation, let us decompose the metric \eqref{gmunu} in the following way:
\beq
g_{\mu\nu}=\eta_{\mu\nu}+h^{\textrm{(M)}}_{\mu\nu}+h^{\textrm{(D)}}_{\mu\nu}+\mathcal{O}\left(r^{-(n+2)}\right)\,.
\eeq
Here $\eta_{\mu\nu}$ is flat space in cylindrical coordinates
\beq\label{cyl}
\eta_{\mu\nu} dx^\mu dx^\nu=-dt^2+dz^2+dr^2+r^2(d\theta^2+\sin^2\theta d\Omega^2_{(n)})\,,
\eeq
and $h_{\mu\nu}^{\textrm{(M)}}$, which is the correction to the flat metric in the {\bf (1st far)} step, includes two types of contributions; There is a $1/R$ change of coordinates of flat space such that $r=0$ corresponds to a circle, which are the terms not multiplying any $r_0$ in \eqref{aasy}. The other type of terms, which are the ones multiplied by $r_0^n$, correspond to the linearized gravitational field sourced by the monopole blackfold stress-energy tensor.

$h^{\textrm{(D)}}_{\mu\nu}$ are the order $r^{-(n+1)}$ terms in the expansion. All these terms have
\beq
h^{\textrm{(D)}}_{\mu\nu}\propto \frac{r_0^{n+2}}{R} \frac{\cos\theta}{r^{n+1}}\,,
\eeq
which is the field of a dipole source located at $r=0$ in the spacetime \eqref{cyl}\footnote{Note that the field of the dipole is insensitive to the fact that $r=0$ is now a circle of curvature $1/R$. This is because the metric \eqref{gmunu} is valid only up to $1/R^2$ corrections. Being the dipole source already a $1/R$ effect, accounting for the fact that the dipole source sits on a circle of curvature $1/R$ is a $1/R^2$ effect.}, namely
\beq
\Delta h^{\textrm{(D)}}_{\mu\nu}\propto \frac{r_0^{n+2}}{R}\partial_{(r\cos\theta)}\delta^{n+2}(r)\,,
\eeq
where $\partial_{(r\cos\theta)}$ points outwards from the worldvolume, such that $\partial_{(r\cos\theta)}\,r=\cos\theta$. The metric \eqref{gmunu} at order $r^{-(n+1)}$ is the linearized solution of
\beq
G_{\mu\nu}[g]=8\pi G\left(B_{\mu\nu}\delta^{n+2}(r)-d_{\mu\nu}{}^{(r\cos\theta)}\partial_{(r\cos\theta)}\delta^{n+2}(r)\right)\,.
\eeq
The goal of this section is to determine $d_{\mu\nu}{}^{(r\cos\theta)}$, which is the dipole source to the far field, and because it is proportional to $1/R$, it is induced by the bending of the monopole source. To find it, we note that in TT gauge, the Einstein equations for $h^{\textrm{(D)}}_{\mu\nu}$ linearize to
\beq
\Delta \bar{h}^{\textrm{(D)}}_{\mu\nu}=16\pi G\, d_{\mu\nu}{}^{(r\cos\theta)}\partial_{(r\cos\theta)}\delta^{n+2}(r)\,,
\eeq
for
\beq
\bar{h}^{\textrm{(D)}}_{\mu\nu}=h^{\textrm{(D)}}_{\mu\nu}-\frac{h^{\textrm{(D)}}}{2}\eta_{\mu\nu}\,,\quad\quad
h^{\textrm{(D)}}=h^{\textrm{(D)}}_{\mu\nu}\eta^{\mu\nu}\,,
\eeq
such that\footnote{We used that in  $D$ spacetime dimensions
$ \Delta_{(D-1)} r^{-(D-3)}  = - (D-3) \Omega_{(D-2)} \delta^{D-1}(r)\,. $}
\beq
\bar{h}^{\textrm{(D)}}_{\mu\nu}=\frac{16\pi G}{\Omega_{n+1}}\frac{\cos\theta}{r^{n+1}}d_{\mu\nu}{}^{(r\cos\theta)}\,.
\eeq
It has been argued in Sec.~\ref{bf2} that $d_{\mu\nu}{}^{(r\cos\theta)}$ has $\mu\nu$ indices parallel to the worldvolume, which is located at $r=0$. $\mu$ and $\nu$ can thus only be $t$ or $z$. This implies that in TT gauge
\beq
h^{\textrm{(D)}}_{rr}=\frac{h^{\textrm{(D)}}_{\Omega\Omega}}{r^2}\,,\quad\quad h^{\textrm{(D)}}_{tt}-h^{\textrm{(D)}}_{zz}=n\,h^{\textrm{(D)}}_{rr}\,,
\eeq
which is satisfied for
\begin{equation}
k_1=\frac{2n}{1-n}\left(k_2-2(n+1)\xi(n)\right)\,,
\end{equation}
which, of course, is a gauge choice. After this gauge fixing, it is straightforward to read $d_{\mu\nu}{}^{(r\cos\theta)}$ from $\bar{h}^{\textrm{(D)}}_{\mu\nu}$. For the sake of interpretation it is convenient to first redefine\footnote{See Eq.~\eqref{FreeEnergy} for a motivation for this.},
\beq
k_2=(n+3)\,\xi(n)+\frac{(n-1)}{2n}\,\tilde{k}_2\,.
\eeq
The dipole contribution is found to be
\begin{subequations}
\beq
d_{tt}{}^{(r\cos\theta)}=-\frac{\Omega_{(n+1)}r_0^n}{16\pi G}\frac{r_0^2}{R}(n^2+3n+4)\xi(n)-\tilde{k}_2\frac{r_0^2}{R}\, B_{tt}\,,
\eeq
\beq
d_{tz}{}^{(r\cos\theta)}=-\tilde{k}_2\frac{r_0^2}{R}\, B_{tz}\,,
\eeq
\beq
d_{zz}{}^{(r\cos\theta)}=\frac{\Omega_{(n+1)}r_0^n}{16\pi G}\frac{r_0^2}{R}(3n+4)\xi(n)\,,
\eeq
\end{subequations}
where $B_{ab}$ in these expressions are \eqref{Tmunu} and are evaluated at equilibrium \eqref{leadingequilrings}. We remind the reader that these expressions are valid for $n>2$.

Some of these terms are not gauge invariant, as they depend on $\tilde{k}_2$. This is the expected `extra symmetry 2'  ambiguity in the dipole under changes of the representative worldvolume surface, see Sec.~\ref{bf2}. Indeed, $\tilde{k}_2\rightarrow\tilde{k}_2+\delta$ picks a  worldvolume outwards by $\delta\, r_0^2/R$. $d_{zz}{}^{(r\cos\theta)}$ being unambiguously defined fits nicely with the fact that the equilibrium condition for the ring is precisely $B_{zz}=0$, which renders this symmetry \eqref{symm2} trivial for this component of the dipole.

\subsection{More general backgrounds}
The calculation of Sec.~\ref{CalculationDipole} can be generalized to black strings lying on flat submanifolds with a more general extrinsic curvature than just non-vanishing $K_{zz}{}^{(r\cos\theta)}$. Ref.~\cite{Caldarelli:2008pz} studied the gravitational field of a black string lying on the $r=0$ submanifold of 
\beq
\begin{split}
ds^2=&-\left(1+C_t\, \frac{2r\cos\theta}{R}\right)dt^2+\left(1+C_z\, \frac{2r\cos\theta}{R}\right)dz^2\\&+\left(1-\frac{C_t+C_z}{n}\frac{2r\cos\theta}{R}\right)\left[dr^2+r^2(d\theta^2+\sin^2\theta\, d\Omega_{(n)}^2)\right]+\mathcal{O}\left((r\cos\theta)^2\right)\,.
\end{split}\label{BackAdS}\eeq
Note that the metric induced on the worldvolume is flat, and the non-vanishing components of its extrinsic curvature read
\beq
K_{tt}{}^{(r\cos\theta)}=\frac{C_t}{R}\,,\quad\quad K_{zz}{}^{(r\cos\theta)}=-\frac{C_z}{R}\,.
\eeq
This family of worldvolumes can be used to perturbatively study black rings in (A)dS space or in Schwarzschild-Tangherlini black hole backgrounds (in which case one is studying Black Saturns in a perturbative regime).

The equilibrium condition for the boosted string lying on $r=0$ solving \eqref{BFeoms} reads
\beq
\sinh^2\alpha=\frac{C_z+(n+1)C_t}{n(C_z-C_t)}\,,
\label{GeneralBoost}\eeq
and the linear blackfold stress-energy tensor takes the form \eqref{Tmunu} with the boost \eqref{GeneralBoost}. A calculation very similar to the one explained in Sec.~\ref{CalculationDipole} on the approximate solution found in \cite{Caldarelli:2008pz} reveals the induced dipole induced by such extrinsic curvature. To write it down in a compact way it is useful to first consider
\beq
K_{ab}{}^\rho d^{ab}{}_\rho=\frac{\Omega_{(n+1)}r_0^n}{16\pi G}\frac{r_0^2}{R}\left[-(3n+4)(C_t^2+C_z^2)-2(n^2+3n+4)C_t C_z\right]\xi(n)\,,
\label{FreeEnergy}\eeq
which is $\tilde{k}_2-$invariant thanks to the leading extrinsic equation of motion, $K_{ab}{}^\rho B^{ab}=0$. The symmetry of this expression under $C_t - C_z$ exchange motivates a parametrization of the gauge ambiguity\footnote{Here, as in Eq.~\eqref{Expansionc}, $k_2$ is defined as the coefficient of the $r_0^{2n+2}/r^{2n+1}$ term in the large $r$ expansion of the $c(r)$ function, appearing in 
Eq.~(B.15) in \cite{Caldarelli:2008pz}.} such that this symmetry is apparent in the $d^{ab\mu}$ object,
\beq
k_2=\frac{\left[(n+1)C_t+(n+3)C_z\right]\left(2C_z+(n+2)C_t\right)}{2(C_z-C_t)}\,\xi(n)+\frac{(n-1)\left[C_z+(n+1)C_t\right]}{2n(C_z-C_t)}\,\tilde{k}_2\,.
\eeq
The induced dipole of a black string on this submanifold reads
\begin{subequations}
\beq
d_{tt}{}^{(r\cos\theta)}=\frac{\Omega_{(n+1)}r_0^n}{16\pi G}\frac{r_0^2}{R}\left[-C_t(3n+4)-C_z(n^2+3n+4)\right]\xi(n)-\tilde{k}_2\frac{r_0^2}{R}\, B_{tt}\,,
\eeq
\beq
d_{tz}{}^{(r\cos\theta)}=-\tilde{k}_2\frac{r_0^2}{R}\, B_{tz}\,,
\eeq
\beq
d_{zz}{}^{(r\cos\theta)}=\frac{\Omega_{(n+1)}r_0^n}{16\pi G}\frac{r_0^2}{R}\left[C_z(3n+4)+C_t(n^2+3n+4)\right]\xi(n)-\tilde{k}_2 \frac{r_0^2}{R}\,B_{zz}\,.
\eeq
\label{gendipole}\end{subequations}
\subsection{The Young Modulus of Black Strings}
The study of how strains induce stresses is the subject of elasticity theory. It is well known from elementary elasticity theory \cite{Landau:1959te} that the bending of an elastic rod induces a stress on it that has opposite signs on the inner and outer side. This is so because under bending, the inner side is compressed and the outer side is stretched. Thus, in classical elasticity theory bent rods develop dipoles of stress.

The geometric object capturing how strain varies on directions transverse to a bent rod is the extrinsic curvature since it can be written 
\beq
K_{\mu\nu}{}^\rho=-\gamma_\mu{}^\kappa\,\gamma_\nu{}^\sigma\,\frac{1}{2}\pounds_{n_\rho} \gamma_{\kappa\sigma}\,,
\eeq
where $\gamma_{\mu\nu}=u^a_\mu\, u^b_\nu\,\gamma_{ab}$ projects on the worldvolume.

Black strings also develop dipoles when bent, exhibiting elastic behavior. In the linear (Hookean) regime, the response coefficient relating stress and strain in non-relativistic elasticity theory is the Young modulus. The black strings we have studied are in such linear regime, as their dipole is a small deformation, of order $1/R$. 

To characterize the elastic behavior of black strings we need to introduce a relativistic generalization of the Young modulus. Let us start by going back to classical elasticity theory in flat space, in which a bent rod develops, in the case $C_t=0$ and $C_z=1$, the stress
\beq
T_{zz}=Y\frac{r\cos\theta}{R}\,,
\eeq
where $Y$ is the Young modulus. If the rod has, to first approximation, a circular cross-section of radius $r_0$, the dipole of stress reads
\beq
d_{zz}{}^{(r\cos\theta)}=\int  T_{zz}\,r\cos\theta\left(r^{n+1}\sin^n\theta\, dr\, d\theta\, d\Omega_{(n)}\right)=\frac{r_0^{n+4}}{(n+2)(n+4)}\frac{\Omega_{(n+1)}}{\Omega_{(n)}}\left(\frac{Y}{R}\right)\,.
\eeq

Motivated by this we now define the Young modulus of black strings through the formula
\beq
d_{ab}{}^\rho=\left[-\frac{r_0^{n+4}}{(n+2)(n+4)}\frac{\Omega_{(n+1)}}{\Omega_{(n)}}\right]Y_{ab}{}^{cd}K_{cd}{}^\rho\,.
\eeq
For relativistic matter we need to use a tensor, accounting for necessary anisotropy of the worldvolume directions: one of the $a, b$ directions is timelike and the rest are spacelike.

It is not the purpose of this work to carry out a deep study of the properties of the relativistic Young modulus that we just defined, $Y_{ab}{}^{cd}$. We note, however, that it should display the `extra symmetry 2' ambiguity that $d_{ab}{}^\rho$ enjoys. By construction, for the result \eqref{gendipole} $\tilde{k}_2=0$ yields
\beq
\left.Y^{ttzz}=Y^{zztt}\right|_{\tilde{k}_2=0}\,,
\eeq
which is a desirable property of such tensors in non-relativistic anisotropic media.

We close this section by collecting the measured components of $Y^{abcd}$ from \eqref{gendipole} at $\tilde{k}_2=0$:
\beq
\begin{split}
Y^{tttt}=&Y^{zzzz}=\frac{\Omega_{(n)}(n+2)(n+4)}{16\pi G\, r_0^2}(3n+4)\xi(n)\,,\quad\quad Y^{tztt}=Y^{tzzz}=0\,,\\
 &Y^{ttzz}=Y^{zztt}=-\frac{\Omega_{(n)}(n+2)(n+4)}{16\pi G\, r_0^2}(n^2+3n+4)\xi(n)\,.
\end{split}\label{YM}\eeq
The addition of flat directions to \eqref{gmunu} or the corresponding case of \eqref{BackAdS} does not threaten the solutions to Einstein equations we have been considering in this section, and does not change the final result \eqref{YM}. Thus, much like the equation of state $\varepsilon=-(n+1)P$ is a property only of the codimension of black $p-$branes, so is the Young modulus, and expressions \eqref{YM} are also valid for $p>1$. Also, because calculations in this section are linear in $1/R$, one can include the effect of a more general extrinsic curvature by just adding the effects of different components up; Considerations in this section are also valid, for example, for black tori \cite{Emparan:2009vd}, in which the $\rho$ index in $K_{ab}{}^\rho$ may not be aligned in all $ab$ components of the extrinsic curvature.
\subsection{Dipole corrections to the equilibrium of thin black rings}
One of the main applications of the blackfold approach has been the construction of approximate black hole solutions. The dipole corrections to the stress-energy tensor of black branes that we have found modify the effective blackfold equations of motion, and can be used to compute next to leading order contributions to these approximate solutions. In this subsection we consider corrections to the approximate construction of thin black rings in flat space.

The relevant blackfold equation for stationary black rings is the extrinsic equation, which is the projection of \eqref{probeapprox} onto directions orthogonal to the worldvolume. It reads
\beq
B^{ab}
K_{ab}{}^\rho =0\,.
\label{leadingeq}\eeq

In flat space, where in the thin limit black rings live on flat submanifolds with $K_{zz}{}^\rho$, these equations become, as discussed around Eq.~\eqref{leadingequilrings},
\beq
B_{zz}=0\,
\eeq

We shall consider corrections to homogeneous, stationary configurations in backgrounds of the type \eqref{BackAdS}. In these backgrounds, the only non-vanishing Christoffel symbols at $r=0$ (apart from the usual coordinate pathology on the transverse sphere) have two parallel indices to the worldvolume:
\beq
u_a^\nu \,u_b^\rho\,\Gamma^{\mu}_{\nu\rho}=K_{ab}{}^{\mu}\,,\quad\quad \textrm{and}\quad\quad u^a_\mu\, u_b^\nu\,\Gamma^{\mu}_{\nu\rho}=- K^{a}{}_{b\rho}\,.
\eeq
These, together with homogeneity, imply
\beq
\nabla_c d^{ab\mu}=-u_d^\mu K_{c}{}^{d}{}_{\rho}d^{ab\rho}\,,\quad\quad
\nabla_c j^{a\mu\nu}=-u_d^\mu K_{c}{}^{d}{}_{\rho}j^{a\rho\nu}-u_d^\nu K_{c}{}^{d}{}_{\rho}j^{a\mu\rho}\,.
\eeq
As has been stressed, this includes black rings in (A)dS and Black Saturns \cite{Elvang:2007rd} with a static central black hole, but excludes other cases, as that of Black Rings in Taub-NUT \cite{Camps:2008hb}.

Under these assumptions, the only non-trivial equation of motion is the extrinsic one \eqref{ext1}, which reduces to
\beq
 m^{ab}K_{ab}{}^\mu-{\perp^\mu}_{\sigma}u_a^\nu S^{\lambda\rho a} R^\sigma{}_{\nu\lambda\rho}=0\,.
\eeq
In terms of $B^{ab}$, $j^{a\mu\nu}$ and $d^{ab\rho}$ these become
\beq
\left(B^{ab}+d^{ca}{}_\rho K_c{}^{b\rho}\right)K_{ab}{}^\mu -{\perp^\mu}_{\sigma}u_a^\nu \left(\frac{1}{2}j^{a\lambda\rho}-d^{ab\lambda}u_b^\rho\right)R^\sigma{}_{\nu\lambda\rho}=0\,.
\eeq
We will now use this formula to draw some predictions, but it is worth noticing that we will not have complete predictability. The reason is that, because the dipole is an induced effect, the $d^{ab\rho}$ corrections that are derived from it are two orders away from the leading order in $r_0/R$, instead of just one, and a complete accounting of effects at this order would require extending the calculations of \cite{Emparan:2007wm, Caldarelli:2008pz} to one order beyond.

For rings in flat space or (A)dS there are no intrinsic angular momentum corrections to \eqref{leadingeq}. This is expected, as these should be insensitive to the orientation of the intrinsic angular momentum (which is in a plane transverse to that of the ring), and thus should be a $j^2$ contribution. When $j^{a\mu\nu}=0$, the leading correction for rings in flat space to \eqref{leadingeq} is the dipole correction
\beq
B_{zz}=-d^{zz(r\cos\theta)}K^z{}_{z(r\cos\theta)}\,,
\eeq
which implies that the critical boost is
\beq
\sinh^2\alpha=\frac{1}{n}+\frac{r_0^2}{R^2}\frac{3n+4}{n}\xi(n)\,.
\label{correquilibrium}\eeq
In the large $n$ limit this becomes
\beq
\sinh^2\alpha=\frac{1}{n}+\frac{r_0^2}{R^2}\left(\frac{3}{n^2}+\dots\right)+\dots\,,
\eeq
highlighting the fact that, at large $n$, the corrections to the single pole account of this type of black holes are further suppressed in $1/n$. Arguments along these lines have been given in the past \cite{Caldarelli:2008mv} stating that this supression is due to the shorter range of the gravitational interaction at large $n$.

The correction term in \eqref{correquilibrium} is two powers in $r_0/R$ away from the leading contribution. This prevents us from computing the conserved charges and thermodynamic properties of these black rings, as we do not have enough data to carry this out; a naive calculation of these properties to the order at which the result \eqref{correquilibrium} is relevant, that is to $(r_0/R)^2$, gives $\tilde{k}_2-$dependent charges, which is unphysical. Only the next order computation in the MAE will cancel the gauge dependence by introducing the unknown ambiguous part in $B_{ab}$, which by Eq.~\eqref{symm2} will be
\beq
B_{ab}\rightarrow B_{ab}+\left(B_{ab}K^{c}{}_{c\rho}+2B^{c}{}_{\left(a\right.}K_{\left.b\right)c\rho}\right)\epsilon^\rho\,,
\label{BabSymm2BR}\eeq
and is of order $(r_0/R)^2$. 

Note, however, that the leading equilibrium condition of black rings in flat space, $B_{zz}=0$, implies $\delta_2 B_{zz}=0$. One then expects Eq.~\eqref{correquilibrium} to hold to order $(r_0/R)^2$, and we trust this equation to the order we have written it. This is not the case for rings in the more general backgrounds \eqref{BackAdS}, and this is why we do not write down a corrected equilibrium condition for \eqref{GeneralBoost}. In conclusion, we cannot at this point predict corrections to the conserved charges of the black rings at order $(r_0/R)^2$. However, we would like to stress that at order $r_0/R$ we do have a prediction, namely that the black rings do not receive corrections for $n>2$.

\section{Corrections for doubly-spinning Myers-Perry black holes} \label{dsbh}
Myers-Perry (MP) black holes exhibit ultra-spinning regimes where the horizon pancakes along one of the planes of rotation \cite{Emparan:2003sy}, a behavior which was later realized also to be shown by higher-dimensional Kerr-(A)dS black holes \cite{Armas:2010hz}. Both of these cases were recently captured within the blackfold framework \cite{Emparan:2009vd},\cite{Armas:2010hz}. The analysis of references \cite{Emparan:2003sy} and \cite{Armas:2010hz} consists in focusing near the axis of rotation and taking a limit such that the horizon looks like a boosted Schwarzschild membrane. On the other hand, having been able to describe this particular limit using the blackfold approach implies that the entire horizon must locally have the geometry of a boosted Schwarzschild membrane. This has not been considered in the literature so far and in App.~\ref{mpsingle} we fill this gap by showing precisely how a regular ultra-spinning limit can be taken everywhere over the horizon.

This section begins with a demonstration of the existence of a similar regime for MP black holes with one non-zero transverse angular momentum. By taking the limit in which the horizon flattens out along one of the planes of rotation and approaching any point on the horizon it is shown that the horizon geometry is locally that of a boosted MP membrane. The extended blackfold formalism presented in Sec.~\ref{bf2} should be able to capture this behavior and, in fact, by reducing Eqs.~\eqref{jc}-\eqref{bound2} to the case under consideration together with a detailed analysis of the thermodynamic properties of such blackfold geometry, this is shown to be the case. The same type of analysis can be carried out for higher dimensional Kerr-(A)dS black holes and it is presented in App.~\ref{kerrads}.

\subsection{Refined ultra-spinning limit} \label{bflimit}

Consider the metric of a MP black hole with two angular momenta in $n+5$ dimensions \cite{Myers:1986un}
\begin{eqnarray}
\label{MP2ang}
ds^2 &=& -dt^2 + \sum_{i=1}^2 \Big[ a_i^2 d\mu_i^2 + (r^2+a_i^2) \mu_i^2 d\phi_i^2 \Big] + \frac{\mu}{r^{n+2} \Pi F} (dt- \sum_{i=1}^2 a_i \mu_i^2 d\phi_i^2 )^2 \nn \\ && + \frac{\Pi F}{\Pi - \frac{\mu}{r^{n+2}}  } dr^2 + r^2 \Big[ d\theta^2 + \cos^2 \theta ( d\psi^2 + \cos^2 \psi d\Omega_{(n-1)}^2 ) \Big]\spa
\end{eqnarray}
where,
\begin{equation} \label{mu12}
\mu_1 = \sin \theta ,\quad\quad \mu_2 = \cos \theta \sin \psi~,
\end{equation}
\begin{equation}
\Pi = \prod_{i=1}^2 (1 + \frac{a_i^2}{r^2}) ,\quad\quad F = 1 - \sum_{i=1}^2 \frac{a_i^2 \mu_i^2}{r^2 + a_i^2} ~.
\end{equation}
The event horizon is located at $r=r_+$ with $r_+$ being the largest positive real root of $r^{n+2} \Pi - \mu = 0$. For clarity of notation, in what follows we label $a_1$ and $a_2$ as $a_1\equiv a$ and $a_2\equiv b$. The aim of this section is to show that there exists an allowed region of parameters where, near the horizon, the metric \eqref{MP2ang}  looks locally like a boosted MP membrane.  To this end, we take the ultra-spinning limit in the first angular momentum parametrized by $a$, that is, $a \gg r_+$.  Since we want to capture the dynamics on the transverse plane $b$ is kept finite such that $a \gg b$. Within such restricted phase space the blackfold looks much like the one encountered in the singly-spinning case\footnote{See App.~\ref{mpsingle} for details.}: a disc with center at $\theta=0$, radius $a$ and boundary at $\theta=\pi/2$.

The assumption of this hierarchy of scales is sufficient to recover the metric of a black membrane near the center of the disc but not everywhere close to the horizon. Instead, the requirement $r \ll a \cos \theta$ is needed. Under this assumption we find
\begin{equation} \nonumber
\Pi  \simeq \frac{a^2}{r^4} ( r^2 + b^2 ) ,\quad\quad F \simeq 1 -\mu_1^2 - \frac{b^2 \mu_2^2}{r^2 + b^2}
,\quad\quad g_{rr} \simeq \frac{(1-\mu_1^2) (r^2 + b^2) - b^2 \mu_2^2}{r^2 + b^2 - \frac{\tilde{\mu}}{a^2 r^{n-2}}}~,
\end{equation}
\begin{equation}\nonumber
\frac{\mu}{r^{n+2} \Pi F} \simeq \frac{\mu}{a^2 r^{n-2} [(1-\mu_1^2) (r^2 + b^2) - b^2 \mu_2^2]}~.
\end{equation}
Now we introduce the coordinate $\rho_1$ as
\begin{equation}
\rho_1 = a \sin \theta~,
\end{equation}
which can be seen as the radius on the disc. Furthermore, assuming also that $b \ll a \cos \theta$ we obtain
\begin{equation}
\sum_{i=1}^2 a_i^2 d\mu_i^2 + r^2 d\theta^2 + r^2 \cos^2 \theta d\psi^2\simeq d\rho_1^2 + \cos^2 \theta (r^2 + b^2  \cos^2 \psi ) d\psi^2~.
\end{equation}
As we are after a description of the local geometry of the horizon we need to approach it at any point, thus we consider the metric near a fixed angle $\theta_*$ (i.e., at a given radius of the disc $\rho_1=\rho_{1*}$). The necessary requirements become $r \ll a \cos \theta_*$ and $b \ll a \cos \theta_*$. To make contact with the metric of a MP membrane written in its usual form, it is convenient to define
\begin{equation} \nonumber
\tilde{r} = r \cos \theta_*  ,\quad\quad \tilde{b} = b \cos \theta_* ,\quad\quad \tilde{r}_0^n = \frac{ \mu( \cos \theta_*)^{n}}{a^2}~,
\end{equation}
\begin{equation}
\label{sigdel}
\Sigma = \tilde{r}^2 + \tilde{b}^2 \cos^2 \psi ,\quad\quad \Delta = \tilde{r}^2 + \tilde{b}^2 - \frac{\tilde{r}_0^n}{\tilde{r}^{n-2}}~.
\end{equation}
Using these definitions we obtain
\begin{equation}
g_{rr} dr^2 \simeq \frac{\Sigma}{\Delta} d\tilde{r}^2 ,\quad\quad \frac{\mu}{r^{n+2} \Pi F} \simeq \frac{\tilde{r}_0^n}{\cos^2 \theta_* \tilde{r}^{n-2} \Sigma}~.
\end{equation}
Next, we introduce the coordinate $z$ by
\begin{equation}
z = \rho_{1*} \phi_1 = a \sin \theta_* \phi_1~,
\end{equation}
which parametrizes the angular direction on the disc. The metric \eqref{MP2ang} near the horizon is now seen to become:
\begin{eqnarray}
\label{MPmembrane1}
ds^2 &=& - dt^2 + d\rho_1^2 + dz^2 + \frac{\tilde{r}_0^n}{\Sigma \tilde{r}^{n-2}} \Big( \frac{dt}{\cos \theta_*} - \tan \theta_* dz - \tilde{b} \sin^2 \psi d\phi_2 \Big)^2 + \frac{\Sigma}{\Delta} d\tilde{r}^2 + \Sigma d\psi^2 \nn \\ &&
+ ( \tilde{r}^2 + \tilde{b}^2 ) \sin^2 \psi d\phi_2^2 + \tilde{r}^2 \cos^2 \psi d\Omega_{n-1}^2 ~.
\end{eqnarray}
This geometry corresponds to the Myers-Perry membrane,
\begin{eqnarray}
\label{MPmembrane}
ds^2 &=& - dt^2 + d\rho_1^2 + dz^2 + \frac{\tilde{r}_0^n}{\Sigma \tilde{r}^{n-2}} ( dt  - \tilde{b} \sin^2 \psi d\phi_2 )^2 + \frac{\Sigma}{\Delta} d\tilde{r}^2 + \Sigma d\psi^2 \nn \\ &&
+ ( \tilde{r}^2 + \tilde{b}^2 ) \sin^2 \psi d\phi_2^2 + \tilde{r}^2 \cos^2 \psi d\Omega_{n-1}^2 ~,
\end{eqnarray}
boosted along the $z$ direction with the boost
\begin{equation}
\label{tzboost}
\left( \begin{array}{c} t \\ z \end{array} \right) = \left( \begin{array}{cc} \frac{1}{\cos \theta_*} & -\tan \theta_* \\ -\tan \theta_* & \frac{1}{\cos \theta_*}  \end{array} \right) \left( \begin{array}{c} \tilde{t} \\ \tilde{z} \end{array} \right)~.
\end{equation}
Applying the boost \eqref{tzboost} to \eqref{MPmembrane} and removing the tildes from $\tilde{t}$ and $\tilde{z}$ leads to the metric \eqref{MPmembrane1}. In turn, \eqref{tzboost} corresponds to the Lorentz boost\footnote{The relation between $\theta_*$ and the rapidity $\eta$ is $\text{tan}\theta_*=\text{sinh}\eta$.}
\begin{equation} \label{lb1}
V = \sin \theta_* = \frac{\rho_1}{a} ,\quad\quad \tilde{\gamma} = \frac{1}{\sqrt{1-V^2}} = \frac{1}{\cos \theta_*} ,\quad\quad \tilde{\gamma} V = \tan \theta_*~.
\end{equation}
In the regime under consideration the MP black hole  given in \eqref{MP2ang} has angular velocity along the $\phi_1$ direction,
\begin{equation} \label{om12}
\Omega_1 \simeq \frac{1}{a}~.
\end{equation}
Thus, in the blackfold description, the ultra-spinning MP black hole with transverse angular momentum is a MP membrane which is rigidly rotating with constant angular velocity,
\begin{equation}
V = \rho_1 \Omega_1~,
\end{equation}
as in the singly-spinning case.

\subsection{Blackfold pole-dipole equations with non-zero transverse angular momentum}\label{bfnz}
As seen above, there exists a limit in which the MP black hole can be locally seen as a black membrane, hence, it should be possible to describe it using the formalism of Sec.~\ref{bf2}. Here, we construct such blackfold geometry by solving Eqs.\eqref{jc}-\eqref{bound2} for a disc-like topology with internal spin current.

Consider $n+5$-dimensional flat spacetime with metric
\begin{equation} \label{ind1}
ds^2 = -dt^2 + d\rho_1^2 + \rho_1^2 d\phi_1^2 + ds_\perp^2+\sum_{i=1}^{n} dx_i^2~,
\end{equation}
where $ds_{\perp}^2$ is the metric on the transverse 2-plane written in the form
\begin{equation} \label{dsperp}
ds^2_\perp=d\rho_2^2+\rho_2^2d\phi_2^2~.
\end{equation}
In this we embed a membrane as
\begin{equation} \label{emb1}
t= \sigma^0 , \quad\quad \rho_1 = \sigma^1 ,  \quad\quad \phi_1= \sigma^2 ,  \quad\quad \rho_2=0 ,  \quad\quad x_i = 0~,
\end{equation}
which gives rise to the induced metric
\begin{equation}\label{ind2}
\gamma_{ab} d\sigma^a d\sigma^b = - dt^2 + d\rho_1^2 + \rho_1^2 d\phi_1^2~.
\end{equation}
The stress-energy tensor of the boosted MP membrane can be read off from \eqref{MPmembrane} and takes the perfect fluid form \eqref{emfluid}, with energy density and pressure given by
\begin{equation}
\varepsilon =  \frac{\Omega_{(n+1)}}{16\pi G} (n+1)\tilde{r}_0^n , \quad\quad P = -  \frac{\Omega_{(n+1)}}{16\pi G}\tilde{r}_0^n ~.
\end{equation}
Since we are trying to construct a black hole solution with a horizon Killing vector field of the form
\beq\label{kvf}
\textbf{k}=\frac{\partial}{\partial t} + \Omega_1\frac{\partial}{\partial\phi_{1}} +  \Omega_2 \frac{\partial}{\partial \phi_2}~,
\eeq
with constant $\Omega_1$ and $\Omega_2$\footnote{The constancy of $\Omega_1$ and $\Omega_2$ need not be imposed. In fact, it is a consequence of the requirement of stationarity. A derivation of these conditions can be accomplished and it will be presented elsewhere.}, then, the fluid velocity must be of the form $u^{a}=k^{a}/|k|$, with non-vanishing components 
\begin{equation} \label{bbf}
u^t = ~\tilde{\gamma} ,\quad\quad u^{\rho_1} = 0 , \quad\quad u^{\phi_1} = \tilde{\gamma} \Omega_1 ,\quad\quad \tilde{\gamma} = \frac{1}{\sqrt{1-\rho_1^2 \Omega_1^2}}~.
\end{equation}
Moreover, the boosted MP membrane spin current can also be obtained from \eqref{MPmembrane} and in this coordinate system simply reads
\begin{equation} \label{Cur}
j^{a\nu\rho} = \frac{\Omega_{(n+1)}}{8\pi G } \frac{ \tilde{b} \tilde{r}_0^nu^a}{\rho_2}\delta_{\rho_2}^{\nu}\delta_{\phi_2}^{\rho}~.
\end{equation}
Since the embedding \eqref{emb1} is completely flat we must have ${K_{ab}}^\rho = 0$. According to the considerations of Sec.~\ref{YMandRings} this immediately implies $d^{ab\rho}=0$. Physically, this is so because there is no bending taking place and hence no elastic forces playing a role. Also, there is no extra contribution to the boundary stress-energy tensor from \eqref{MPmembrane} leading to a vanishing $B^{\mu\nu a}$ everywhere. This is most likely due to the fact that the blackfold boundary is regular, containing no extra stress-energy sources. Finally, we need to know the transverse angular velocity $\Omega_2$ as a function of $\tilde{r}_0^n$ and $\tilde{b}$. This can be seen as an additional equation of state and is given by
\begin{equation} \label{tom2}
\Omega_2 (\tilde{r}_0^n,\tilde{b}) =\frac{1}{\tilde{\gamma}} \frac{\tilde{b}}{\tilde{r}_+(\tilde{r}_0^n,\tilde{b})^2+\tilde{b}^2} ~,
\end{equation}
where $\tilde{r}_+(\tilde{\mu},b)$ is found as the highest real root of $\Delta(\tilde{r})$, see \eqref{sigdel}, thus,
\begin{equation}
\label{br0mu}
 \tilde{r}_+^2 + \tilde{b}^2 = \frac{\tilde{r}_0^n}{\tilde{r}_+^{n-2}}~.
\end{equation}
With this in hand, the pole-dipole blackfold equations \eqref{jc}-\eqref{bound2}, reduce in this case to:
\begin{equation} \label{bfmppd}
B^{ab}{K_{ab}}^{\rho}=0\spa D_a B^{ab} = 0 \spa \perp_{\lambda}^{\mu}\perp_{\rho}^{\mu}\nabla_a j^{a\nu\rho} = 0 \spa B^{ab}u^{\mu}_{a}n_b|_{\partial\mathcal{W}_{3}}=0 \spa j^{a\mu\nu}n_{a}|_{\partial\mathcal{W}_{3}}=0~.
\end{equation}
The first and last equations in \eqref{bfmppd} are trivially satisfied due to the vanishing of the extrinsic curvature tensor and to the vanishing of the $\rho_{1}$ component of the fluid velocity $u^{a}$ respectively.  The boundary condition $B^{ab}u^{\mu}_{a}n_b|_{\partial\mathcal{W}_{3}}=0$ requires that, at the boundary, the fluid must be moving with the speed of light, i.e., $\rho_{1}|_{\partial\mathcal{W}_3}=\Omega_{1}^{-1}$. 

Furthermore, assuming that $\epsilon$ and $P$ only depend on $\rho_1$, we find from the conservation of the stress-energy tensor
\begin{equation}
D_a B^{ab} = (\epsilon + P) \dot{u}^b + \partial^b P~,
\end{equation}
which is solved by,
\begin{equation}\label{musol}
\tilde{r}_0 (\rho_1) = \tilde{r}_0 (0) \sqrt{1-\rho_{1}^2 \Omega_1^2}~.
\end{equation}
Since $ j^{a\nu\rho} $ is independent of $t$ and $\phi_1$, the conservation equation $\perp_{\lambda}^{\mu}\perp_{\rho}^{\mu}\nabla_a j^{a\nu\rho} = 0$ is trivially obeyed, $i.e.$, it does not constrain $\tilde{b}$ as a function of $\rho_1$. According to \eqref{kvf} we now use the constancy of $\Omega_2$ over the blackfold. As a function of $\rho_1$ we see from Eq.~\eqref{musol} that $\tilde{r}_0^n$ is proportional to $\gamma(\rho_1)^{-n}$. Using Eq.~\eqref{tom2} and Eq.~\eqref{br0mu} we see that both $\tilde{b}$ and $\tilde{r}_0^n$ should be proportional to $\gamma(\rho_1)^{-1}$. Thus, we find
\begin{equation}
\tilde{b}(\rho_1) = \tilde{b}(0) \sqrt{1-\rho_1^2 \Omega_1^2}~. 
\end{equation}
In order to show that this blackfold solution corresponds indeed to the limit taken above in the doubly-spinning MP metric \eqref{MP2ang} we proceed by computing its thermodynamic properties.

\subsection{Thermodynamic quantities}
The thermodynamic properties for the analytic solution given in \eqref{MP2ang} in the ultra-spinning regime can be obtained from reference \cite{Gibbons:2004ai} and read,
\beq \label{set1}
M=\frac{\Omega_{(n+3)}}{16\pi G}\mu (n+3),\quad\quad J_{1}=\frac{\Omega_{(n+3)}}{8\pi G}\mu a,\quad\quad J_{2}=\frac{\Omega_{(n+3)}}{8\pi G}\mu b ~,
\eeq
with angular velocities
\beq\label{and12}
\Omega_{1}=\frac{1}{a_1},\quad\quad\Omega_{2}=\frac{b}{r_{+}^2+b^2}~,
\eeq
while the temperature and entropy are given by,
\beq \label{set2}
S=\frac{\Omega_{(n+3)}}{4 G}\frac{a^2 \mu}{r_+},\quad\quad T=\frac{1}{4\pi r_{+}}\left( n - \frac{2b^2}{r_{+}^2+b^2} \right)~.
\eeq
We want to show that this is correctly reproduced from the blackfold description. To this aim, we use the expressions \eqref{m}-\eqref{jt} together with \eqref{emfluid},\eqref{Cur} and find,
\beq \label{set3}
M=\frac{\Omega_{(n+3)}}{16\pi G}\frac{\tilde{r}_0^n(0)}{\Omega_{1}^2}(n+3),\quad\quad J^{1}=\frac{\Omega_{(n+3)}}{8\pi G}\frac{\tilde{r}_0^n(0)}{\Omega_{1}^2}\frac{1}{\Omega_1},\quad\quad J_{\perp}^{2}=\frac{\Omega_{(n+3)}}{8\pi G}\frac{\tilde{r}_0^n(0)}{\Omega_{1}^2}\tilde{b}(0)~.
\eeq 
We see that we find perfect agreement between the above quantities and those presented in \eqref{set1}-\eqref{and12} if we identify $\tilde{r}_0^n (0)/\Omega_1^2=\mu$, $\Omega_1=a^{-1}$ and $\tilde{b}(0)=b$. To compute the entropy and temperature we use the method described in Sec.~\ref{thermo}. We first start by computing the product $TS$ using the Smarr relation \eqref{smarr}, yielding,
\beq\label{ts1}
TS=\frac{\Omega_{(n+3)}}{16\pi G}\frac{\tilde{r}_0^n(0)}{\Omega_{1}^2}\left(n-\frac{2\tilde{b}(0)^2}{r_{+}^2+\tilde{b}(0)^2}\right)~.
\eeq 
All the quantities given in \eqref{set3} and the product above can be parametrized in terms of $r_+$ and $b(0)$ using Eq.~\eqref{br0mu}. According to Eq.~\eqref{thermosets} we obtain a set of two equations, for which the solution is,
\beq
T=\frac{\tilde{C}}{r_{+}}\left(n-\frac{2\tilde{b}(0)^2}{r_{+}^2+\tilde{b}(0)^2}\right)~.
\eeq
The constant $\tilde{C}$ can be fixed by requiring the right result in the infinitely thin limit, i.e., when $\tilde{b}(0)=0$. This implies $\tilde{C}=1/4\pi$. It is easy to check that this matches with the temperature given in \eqref{set2}. The entropy then follows by using Eq.~\eqref{ts1}. In this way, we have indeed shown that the ultra-spinning MP black hole with one transverse angular momentum is accurately described within the blackfold framework in the pole-dipole approximation.

\section{Conclusions and outlook} \label{cout}
This paper has studied corrections to the leading blackfold approach that appear as dipole contributions to the effective blackfold stress-energy tensor. The consideration of these corrections required the introduction of a new response coefficient, the Young modulus. The physics captured by the Young modulus is of the same type as that of the Love numbers of black holes\footnote{See Ref.~\cite{Kol:2011vg}, that appeared when this paper was being finished, and with which our Sec.~\ref{YMandRings} tangentially overlaps.}. The calculation of the Young modulus presented here is similar in spirit to the calculations of transport coefficients in the context of the AdS/CFT correspondence\footnote{As a matter of fact, many aspects of the blackfold derivative expansion are reminiscent of the Fluid/Gravity correspondence \cite{Bhattacharyya:2008jc, Baier:2007ix, Hubeny:2011hd}.}. Because the Young modulus is active when there is bending, introducing this type of response coefficient in AdS/CFT would require considering the bending of the brane. There is more to this than non-normalizable fluctuations of AdS; Non-normalizable modes of AdS change the intrinsic curvatures in the boundary, whereas the bending of a brane manifests itself through the extrinsic curvature. This is seen, to leading order, as a dipole deformation of the transverse sphere; for the AdS/CFT of D$3$ branes bending corresponds to deforming the $S^5$ of $\textrm{AdS}_5\times S^{5}$, which is dual to sourcing the scalars of $\mathcal{N}=4$ transforming in the fundamental of the $SO(6)$ R-symmetry group.

In the blackfold approach black holes are seen as probe branes characterized by fluid dynamics and material science concepts: they have an energy density, pressure, viscosity \cite{Camps:2010br}, Young modulus, etc. This should not come as a surprise. We have argued that the blackfold approach is a long wavelength effective theory, and so are material science and fluid dynamics. In these theories, the small scale is an interatomic distance or a thermal wavelength, and in the blackfold approach it is the radius of the transverse sphere. Material science and fluid dynamics are universal long wavelength theories and is natural we recover them in the blackfold limit. Because the fluid behavior enters in corrections to the intrinsic stress-energy tensor ($B_{ab}$) and the material behavior shows up as extrinsic dipole corrections ($d_{ab}{}^\rho$), we can say that blackfolds behave as liquids in worldvolume directions and solids in transverse directions\footnote{We thank Roberto Emparan for suggesting this.}.

As argued in \cite{Vasilic:2007wp}, the dipole contribution to the stress-energy effective tensor suffers an ambiguity, that we parametrized by $\tilde{k}_2$. The origin of this ambiguity is the freedom to choose a representative surface inside the finitely thick black brane as the worldvolume surface. $\tilde{k}_2$ shifts the representative worldvolume by $\epsilon^\rho\partial_\rho=\tilde{k}_2\frac{r_0^2}{R}\partial_{(r\cos\theta)}$. Although the black brane has thickness $r_0$ we are only free to move the representative surface by a much smaller amount, of about $r_0^2/R$. This is so because shifts by larger amounts threaten the hierarchy $B^{\mu\nu\rho}/B^{\mu\nu}=r_0/R=\mathcal{O}_1$, which is crucial for the consistency of the pole-dipole approximation.

In the point particle case there is a natural way to fix this ambiguity, namely by requiring the mass dipole to vanish: This natural representatitve is the center of mass, whose coordinates are defined as
\beq
X^\mu=\frac{1}{m}\int_\Sigma \sqrt{-g}\,d^{D-1} x\, T^{00}\,x^\mu\,.
\eeq
For $p-$branes this choice is not well defined in general, simply because there is more than one worldvolume dipole, $d^{ab\rho}$. While the vanishing of one particular component may be achieved by gauge fixing, the others will not vanish in general.

The monopole contribution to the effective stress-energy tensor also suffers from this ambiguity \eqref{symm2}, but for the monopole this is a $(r_0/R)^2$ contribution \eqref{BabSymm2BR}, and this is why we have not seen it at the order we have worked in the MAE. Doing the MAE to next order would allow us to see this contribution to $B_{ab}$. A natural way of fixing this symmetry would be demanding that the equation of state of $B_{ab}$ remains $\varepsilon=-(n+1)P$.

It will be very interesting to use the analysis we have carried out here to study other blackfold configurations, like those in which one will see a non-zero spin-spin interaction, as in Black Saturns with a central rotating Myers-Perry black hole.

An important generalization of what we have done is the dielectric case we explained at the end of the introduction, instead of only its bending analog. The blackfold approach has been developed for charged black branes \cite{Caldarelli:2010xz, Emparan:2011hg, Grignani:2010xm} and, under the influence of background $p-$form fields, blackfolds will behave dielectrically and paramagnetically, developing charge and magnetic dipoles. In this case, the response coefficients are polarizabilities and susceptibilities.

Finally, it would be interesting to study time-dependent embeddings, from which one could measure a viscosity that would give the time scale of the damping of extrinsic oscillations of black strings. This viscosity is different from those measured in \cite{Camps:2010br}. Using a mixed analysis, in which both types of viscosities (intrinsic and extrinsic) would be active, one could address the very important problem of the stability of thin black rings in a controlled approximation, to an order accounting for the Gregory-Laflamme long-wavelength cutoff. We plan to address this problem in the near future.

\section*{Acknowledgements}
We thank Marco Caldarelli, Roberto Emparan, Veronika Hubeny, Barak Kol, David Mateos, Shiraz Minwalla, Mukund Rangamani and Simon Ross for discussions. Jay would like to thank the Tata Institute for Fundamental Research for hospitality during the completion of a part of this work and to FCT Portugal for the grant SFRH/BD/45893/2008. JC is supported by the STFC Rolling grant ST/G000433/1. TH thanks the Niels Bohr Institute for hospitality. JC, TH and NO would like to thank the Centro de Ciencias de Benasque Pedro Pascual, and JC  would also like to thank the NBI and ICTP for hospitality during various stages of this project. JC acknowledges support of GEOMAPS during his stay at the NBI.
The work of NO is supported in part by the Danish National Research Foundation project ``Black holes and their role in quantum gravity''.

\appendix

\section{Notation}\label{notation}
In this appendix we collect some conventions about notation. This paper deals with submanifolds $\mathcal{W}_{p+1}$ of a $(D=p+n+3)-$dimensional spacetime, described by the mapping functions $X^\mu (\sigma^a)$ from the worldvolume, parametrized by the coordinates $\sigma^a$, to the ambient spacetime, with coordinates $x^{\mu}$.

$g_{\mu\nu}$ is the background metric while $\mathcal{W}_{p+1}$ inherits the metric
\beq
\gamma_{ab}=u_a^\mu\, g_{\mu\nu}\, u_b^\nu\,,\quad\quad u_a^\mu=\partial_a X^\mu\,.
\eeq
Here, $\mu,\nu$ indices are raised and lowered with $g_{\mu\nu}$ and its inverse $g^{\mu\nu}$, and $a,b$ indices with $\gamma_{ab}$ and its inverse $\gamma^{ab}$.

To project any spacetime tensor along tangential directions to the worldvolume we can use $u^a_\mu$, while for directions orthogonal to $\mathcal{W}_{p+1}$ we define the projector,
\beq
\perp_{\mu\nu}=g_{\mu\nu}-u^a_\mu\,\gamma_{ab}\,u^b_\nu\,.
\eeq
A subindex $_\perp$ on a tensor indicates that all $\mu, \nu$ type of indices are orthogonal, e.g.,
\beq
B_\perp^{a\mu}={\perp^\mu}_\nu B_\perp^{a\nu}\,.
\eeq
$j^{a\mu\nu}$ and $d^{ab\mu}$ defined in \eqref{j} and \eqref{d} are $_\perp$-objects, but we do not write the $_\perp$ subindex on them in order to avoid cluttering.

$\nabla_\mu$ is the covariant derivative on the ambient space compatible with $g_{\mu\nu}$ and $\Gamma_{\mu\nu}^\rho$ are its Christoffel symbols. $D_a$ is the intrinsic covariant derivative on $\mathcal{W}_{p+1}$ compatible with $\gamma_{ab}$ and $\left\{\begin{smallmatrix}a\\b\,c\end{smallmatrix}\right\}$ are its Christoffel symbols. Our convention for the Riemann curvature tensor is
\beq
{R^{\mu}}_{\lambda\nu\rho}=\Gamma^{\mu}_{\lambda\rho,\nu}-\Gamma^{\mu}_{\lambda\nu,\rho}+\Gamma^{\mu}_{\sigma\nu}\Gamma^{\sigma}_{\lambda\rho}-\Gamma^{\mu}_{\sigma\rho}{\Gamma^{\sigma}}_{\lambda\nu}\,.
\eeq

The operator $\nabla_a$ is defined to be compatible both with $\gamma_{ab}$ and $g_{\mu\nu}$ such that, for instance,
\beq
\nabla_a v_{\mu}{}^{\nu}{}_b{}^c=u_a^\rho\,\partial_\rho\, v_{\mu}{}^{\nu}{}_b{}^c
-u_a^\rho\,\Gamma_{\rho\mu}^\sigma\, v_{\sigma}{}^{\nu}{}_b{}^c
+u_a^\rho\,\Gamma_{\rho\sigma}^\nu\, v_{\mu}{}^{\sigma}{}_b{}^c
-\left\{\begin{smallmatrix}d\\a\,b\end{smallmatrix}\right\}\, v_{\mu}{}^{\nu}{}_d{}^c
+\left\{\begin{smallmatrix}c\\a\,d\end{smallmatrix}\right\}\, v_{\mu}{}^{\nu}{}_b{}^d\,.
\eeq
For a submanifold tensor, $\nabla_c\, t^{a\dots}{}_{b\dots}=D_c\, t^{a\dots}{}_{b\dots}$. Moreover, the extrinsic curvature of $\mathcal{W}_{p+1}$ can be written as
\beq
\nabla_a u_b^\rho=K_{ab}{}^\rho\,,
\eeq
with $K_{ab}{}^\rho$ being also a $_\perp-$object.

The boundary of the submanifold is described by $\sigma^a=\zeta^{a}(\lambda)$ and normal vector $\hat{n}^{\mu}=u^{\mu}_{a}\hat{n}^{a}$ with unit norm. We introduce the coordinate vectors $v^{a}_{\hat{i}}$ as
\beq
v^{a}_{\hat{i}}=\frac{\partial\zeta^a}{\partial\lambda^{\hat{i}}}~,
\eeq
satisfying the properties $v^{\mu}_{\hat{i}}=u^{\mu}_{a}v^{a}_{\hat{i}}$ and $\hat{n}_{a}v^{a}_{\hat{i}}=0$ such that the induced metric on the boundary takes the form $\hat{h}_{\hat{i}\hat{j}}=\gamma_{ab}(\zeta)v_{\hat{i}}^av_{\hat{j}}^b$. $\nabla_{\hat{i}}$ is the boundary covariant derivative compatible with the metric $\hat{h}_{\hat{i}\hat{j}}$.
\section{Ultra-spinning Myers-Perry black holes revisited} \label{mpsingle}
The purpose of this section is to instructively show how to take the blackfold limit for singly-spinning MP black holes as it was done in Sec.~\ref{bflimit} for its doubly-spinning counterpart. Bearing this in mind, we consider the MP metric with a single angular momentum in $n+5$ dimensions
\begin{equation} \label{mp1}
ds^2 = -dt^2 + \frac{\mu}{r^{n} \Sigma} ( dt - a \sin^2 \theta d\phi )^2 + \frac{\Sigma}{\Delta} dr^2 + \Sigma d\theta^2 + (r^2 + a^2 ) \sin^2 \theta d\phi^2 + r^2 \cos^2 \theta d\Omega_{(n+1)}^2~,
\end{equation}
with
\begin{equation}
\Sigma = r^2 + a^2 \cos^2 \theta ,\quad\quad \Delta = r^2 + a^2 - \frac{\mu}{r^{n}}~. 
\end{equation}
The horizon radius $r_+$ is given by the largest positive real root of $\Delta(r)=0$. The ultra-spinning limit is attained when $r_+ \ll a$ \cite{Emparan:2003sy}. From the definition of $\Delta$ above we see that in this limit $r_+^{n} \simeq \mu/ a^2$. From the $n+1$-sphere metric in \eqref{mp1} we see that the radius of the $(n+1)$-sphere is
\begin{equation}
r_0= r_+ \cos \theta =  \left( \frac{\mu}{a^2} \right)^{\frac{1}{n}} \cos \theta~.
\end{equation}
Hence, the blackfold is a rotating disc of radius $a$ with its center and boundary located at $\theta=0$ and $\theta=\pi /2$ respectively.  We can see this directly from the metric \eqref{mp1} above. Sufficiently close to the horizon, the metric \eqref{mp1} should locally be that of a boosted black membrane. Close to the center of the disc we need $r \ll a$. However, this is not sufficiently close everywhere on the blackfold. We thus introduce the coordinate
\begin{equation}
\rho_1 = a \sin \theta~.
\end{equation}
This can be seen as the radius on the disc. In terms of this we can write the thickness as
\begin{equation} \label{t1}
r_0(\rho_1) = r_+ \sqrt{1-\frac{\rho_1^2}{a^2}} ~.
\end{equation}
For a given point on the disc we should require that the distance scale over which the thickness change is much larger than the thickness of the disc. Thus, we should require
\begin{equation}
r_0 \ll \frac{1}{| r_0'' |}~.
\end{equation}
In terms of the horizon radius and rotation parameter, this requirement becomes:
\begin{equation}
r_+ \ll a \sqrt{1-\frac{\rho_1^2}{a^2}}~.
\end{equation}
This tells us, for each point on the disc, how widely separated the scales must be for the blackfold approximation to be valid. To see the boosted black membrane from the metric \eqref{mp1}, given a radius $\rho_1$, we need
\begin{equation}
r \ll a \sqrt{1-\frac{\rho_1^2}{a^2}}~.
\end{equation}
This implies $r \ll a \cos \theta$.  Now, consider a given point with radius $\rho_1=\rho_{1*}$ at the disc, corresponding to the angle $\theta_*$ with $\rho_{1*} = a \sin \theta_*$. Hence, we require $r \ll a \cos \theta_*$. In order to make a more clear contact with the metric of a Schwarzschild membrane we define
\begin{equation}
r_{0*} = r_+ \cos \theta_*,\quad\quad \tilde{r} = r \cos \theta_*,\quad\quad~z = \rho_{1*} \phi~.
\end{equation}
In the approximation $r \ll a \cos \theta_*$ the metric near $\rho=\rho_{1*}$ becomes:
\begin{equation}
\label{metmpnear}
ds^2 = -dt^2  + d\rho_1^2 +  dz^2  + (1-f) \left( \frac{dt}{\cos \theta^*} - \tan \theta_* dz \right)^2  
+ \frac{d\tilde{r}^2}{f} + \tilde{r}^2 d\Omega^2_{(n+1)} ~,
\end{equation}
with
\begin{equation} \label{rs}
f \equiv 1- \frac{r_{0*}^{n}}{\tilde{r}^{n}}~.
\end{equation}
This corresponds to boosting the static black Schwarzschild membrane
\begin{equation}
\label{staticmembrane}
ds^2 =- dt^2 + dz^2 + (1-f) dt^2 + d\rho_1^2+ \frac{d\tilde{r}^2}{f} + \tilde{r}^2 d\Omega^2_{(n+1)} ~,
\end{equation}
along $z$ with the boost \eqref{tzboost} with corresponding Lorentz boost parameter \eqref{lb1}. The angular velocity in this ultra-spinning regime is also given by \eqref{om12} and hence we see that in the blackfold description the ultra-spinning MP black hole is a black membrane which is rigidly rotating with constant angular velocity given by $V=\rho_1\Omega_1$.

\subsection*{Blackfold equations with zero transverse angular momentum}
To facilitate comparison with the doubly-spinning case of Sec.~\eqref{bfnz} we present here a brief analysis of the blackfold equations in the single-pole approximation. 

We begin by embedding the disc in the background \eqref{ind1} using the mapping functions \eqref{emb1}, leading to the induced metric \eqref{ind2}. Next, we read off the static black membrane stress-energy tensor from the metric \eqref{metmpnear}, which has the form of \eqref{emfluid}. The fluid velocity is given by the pullback of the background Killing vector field
\begin{equation}
\textbf{k}=\frac{\partial}{\partial t}+\Omega_1\frac{\partial}{\partial \phi_1}~,
\end{equation}
which gives rise to the same non-vanishing components as those given in \eqref{bbf}.

Now, we analyze the blackfold equations \eqref{jc}-\eqref{bound2} adapted to the current situation. As the embedding is flat, the extrinsic curvature $K_{ab}{}^\mu$ vanishes and thus the extrinsic equation $T^{ab} K_{ab}{}^\mu = 0$ is trivially satisfied. The remaining non-trivial equations read
\begin{equation}
D_a B^{ab} = 0 ,\quad\quad B^{ab}u^{\mu}_{a}n_b|_{\partial\mathcal{W}_{3}}=0~.
\end{equation}
The conservation of the intrinsic stress-energy tensor, assuming that $\epsilon$ and $P$ only depend on $\rho_1$, implies
\begin{equation}
r_0 (\rho_1) = r_0 (0) \sqrt{1-\rho_1^2 \Omega_1^2}~.
\end{equation}
Furthermore the boundary condition $B^{ab}u^{\mu}_{a}n_b|_{\partial\mathcal{W}_{3}}=0$ is again satisfied provided $\rho_{1}|_{\partial\mathcal{W}_{3}}=\Omega_{1}^{-1}$, hence the disc has a radius of $\rho_{1}=\Omega_{1}^{-1}$ and is moving at the speed of light there. This is the result obtained in \cite{Emparan:2009cs, Emparan:2009vd} and matches the thickness \eqref{t1} upon the identification $r_0(0)=(\mu/a_1^2)^{\frac{1}{n}}$ and $\Omega_1=a_1^{-1}$.

\section{Spin corrections for higher-dimensional Kerr-(A)dS black holes} \label{kerrads}
Here we study the same limiting behavior of higher-dimensional Kerr-(A)dS black holes as in Sec.~\ref{dsbh} for MP black holes and show that it can be described using the blackfold formalism. Due to the similarity between both cases we refer to Sec.~\eqref{dsbh} for a more extensive analysis.

In spheroidal coordinates, the metric of the Kerr-AdS black hole with two angular momenta in even dimensions is given by\footnote{The same analysis also holds in the case of odd dimensions.} \cite{Hawking:1998kw, Gibbons:2004uw}
\beq
\begin{split} \label{kads}
ds^2=&-W\left(1+\frac{r^2}{L^2}\right)dt^2+\frac{\mu}{U}\left(dt-\sum_{i=1}^{2}\frac{a_{i}\mu_{i}^2}{\Xi_{i}}d\phi_{i}\right)^2 + \frac{U}{V-\mu}dr^2 \\
&+\sum_{i=1}^{2}\frac{r^2+a_{i}^2}{\Xi_{i}}\left(d\mu_{i}^2+\mu_{i}^2\left(d\phi_{i}+\frac{\sqrt{\alpha_{i}}}{L}dt\right)^2\right)+r^2\left(\sum_{i=3}^{(n+5)/2}d\mu_{i}^2+\sum_{i=3}^{N}\mu_{i}^2d\phi_{i}^2\right) \\ 
&+\frac{1}{L^2 W(1+\frac{r^2}{L^2})}\left(\sum_{i=1}^{2}\frac{r^2+a_{i}^2}{\Xi_{i}}\mu_{i}d\mu_{i}+r^2\sum_{i=3}^{(n+5)/2}\mu_{i}d\mu_{i}\right)~,
\end{split}
\eeq
where
\beq
\begin{split}
W=1+\sum_{i}^{2}\alpha_{i}&\frac{\mu_{i}^2}{\Xi_{i}},\quad\quad U=r^{n-2}\left(1-\sum_{i=1}^{2}\frac{a_{i}^2\mu_{i}^2}{r^2+a_{i}^2}\right)\prod_{j=1}^{2}(r^2+a_{j}^2), \\
&\quad\quad V=r^{n-2}(1+\frac{r^2}{L^2})\prod_{i=1}^{2}(r^2+a_{i}^2)~,
\end{split}
\eeq
with $\mu_{1}$ and $\mu_2$ as given in Eq.~\eqref{mu12} and
\beq
\frac{U}{V-\mu}= \frac{\left(1-\sum_{i=1}^{2}\frac{a_{i}^2\mu_{i}^2}{r^2+a_{i}^2}\right)}{(1+\frac{r^2}{L^2})-\frac{\mu}{r^{n-2}\prod_{j=1}^{2}(r^2+a_{j}^2)}},\quad\quad \alpha_{i}=\frac{a_{i}^2}{L^2},\quad\quad\Xi_{i}=1-\alpha_{i}~.
\eeq
The horizon $r_{+}$ is given by the largest positive real root of $V(r)-\mu=0$. For clarity of notation we set $a_1\equiv a$ and $a_2\equiv b$. The ultra-spinning limit is attained when $r_{+},b\ll a,L$ with $0\le \alpha_{1}<1$ and $r\ll a,L$ with $r$ finite\footnote{In the case of higher-dimensional Kerr-dS black holes,  the parameter $\alpha_{1}$ is instead constrained by $\alpha_{i}\ge0$.}, in a similar fashion as in the singly-spinning case \cite{Armas:2010hz}. However, for the metric to look locally like a boosted MP membrane, we furthermore need $r\ll \frac{a}{\sqrt{\Xi_{1}}}\text{cos}\theta$ and $b\ll \frac{a}{\sqrt{\Xi_{1}}}\text{cos}\theta$.

Under these assumptions, it is easy to show that the last term in \eqref{kads} is subleading while the functions $V,U,W$ reduce to
\beq
W\to1,\quad\quad U\to r^{n-2}a^2\left(1-\mu_{1}^2-\frac{a_{i}^2\mu_{i}^2}{r^2+a_{i}^2}\right)(r^2+b^2),\quad\quad V\to r^{n-2}a^2(r^2+b^2)~.
\eeq
In order to parametrize the membrane in a convenient way, we introduce the coordinates
\beq
\rho_1=\frac{a}{\sqrt{\Xi_{1}}}\text{sin}\theta ,\quad\quad z=\rho_{*}\phi_{1}~.
\eeq
Then, near a fixed angle $\theta_*$, the metric \eqref{kads} is seen to reduce again to that of a MP membrane \eqref{MPmembrane} but with Lorentz boost,
\beq \label{boost}
V=\text{sin}\theta_{*}=\sqrt{\Xi_{1}}\Omega_1\rho_1,\quad\quad \tilde{\gamma}=\frac{1}{\sqrt{1-\Xi_{1}\rho_1^2\Omega_{1}^2}},\quad\quad \Omega_{1}=\frac{1}{a}~.
\eeq

\subsection*{Blackfold pole-dipole equations with non-zero transverse angular momentum}
We now want to describe the above limiting behavior of the higher-dimensional Kerr-(A)dS black holes using the blackfold approach. It is convenient, in order to highlight the existent 2-planes of the background spacetime, to write the metric of AdS in conformally flat coordinates
\beq
ds^2=-F(\rho)dt^2+H(\rho)^{-1}\left(d\rho_1^2+\rho_1^2d\phi_1^2+ds^2_{\perp}+\sum_{i=1}^{n} dx_i^2\right)\spa\rho^2=\rho_1^2+\rho_2^2+\sum_{i=1}^{n} x_{i}^2~,
\eeq
where $ds_{\perp}^2$ is given by \eqref{dsperp} and
\beq
F(\rho)=\left(\frac{1+\frac{\rho^2}{4L^2}}{1-\frac{\rho^2+}{4L^2}}\right)^{2}, \quad\quad H(\rho)=1-\frac{\rho^2}{4L^2}~.
\eeq
Note that this coordinate system differs from the one used in \eqref{kads}. To see how to translate from one coordinate system to the other see for example \cite{Armas:2010hz}. To embed the membrane in this background we chose the embedding coordinates \eqref{emb1}, which give rise to the induced metric
\begin{equation}
\gamma_{ab} d\sigma^a d\sigma^b = - F(\rho_1)dt^2 + H(\rho_1)^{-1}\left(d\rho_1^2 + \rho_1^2 d\phi_1^2\right)~.
\end{equation}
Again, all extrinsic curvature components vanish since the embedding is flat. The stress-energy tensor is still given by \eqref{emfluid} but now with boost velocities,
\begin{equation} 
u^t = \tilde{\gamma}, \quad\quad u^{\rho_1} = 0 , \quad\quad u^{\phi_1} = \tilde{\gamma} \Omega_1, \quad\quad \tilde{\gamma}=\frac{1}{\sqrt{1-\Xi_1\frac{\rho_1^2}{1-\frac{\rho_1^2}{4L^2}}\Omega_{1}^2}}~.
\end{equation}
Furthermore the spin current in this coordinate system reads
 \begin{equation} 
j^{a\nu\rho}=\frac{\Omega_{(n+1)}}{8\pi G } \frac{\tilde{b} \tilde{r}_0^n H(\rho)}{\rho_2}  u^a\delta_{\rho_2}^{\nu}\delta_{\phi_2}^{\rho}~,
\end{equation}
while the dipole current and $B^{\mu\nu a}$ components vanish. Due to the vanishing of the extrinsic curvature and of all the contractions involving the Riemann tensor in Eqs.~\eqref{jc}-\eqref{bound2}, the blackfold equations reduce to Eqs.~\eqref{bfmppd} as in the flat space case. Solving the bulk equations requires:
\beq \label{solker}
\tilde{r}_0(\rho_1)=\tilde{r}_0(0)\sqrt{1-\Xi_1\frac{\rho_1^2}{1-\frac{\rho_1^2}{4L^2}}\Omega_{1}^2},\quad\quad \tilde{b}(\rho_1)=\tilde{b}(0)\sqrt{1-\Xi_1\frac{\rho_1^2}{1-\frac{\rho_1^2}{4L^2}}\Omega_{1}}~.
\eeq
Moreover the boundary condition $B^{ab}u^{\mu}_{a}n_b|_{\partial\mathcal{W}_{3}}=0$ implies that the disc has a maximum radius given by $\rho_1|_{\partial\mathcal{W}_{3}}=2L(L\Omega_1-\sqrt{L^2\Omega_1^2-1})$.
We note that in the singly-spinning case where $\tilde{b}(0)=0$ the thickness $\tilde{r}_{0}$ obtained in \cite{Armas:2010hz} coincides with the one given in \eqref{solker} and agrees with the analytic solution \eqref{kads} upon the identification $\tilde{r}_0^n(0)/\Omega_1^2=\mu$ and $\Omega_{1}^{-1}=a_{1}$. 

\subsection*{Thermodynamic quantities}
The thermodynamical quantities of the analytic solution \eqref{kads} in the ultra-spinning regime can be obtained from \cite{Gibbons:2004ai} and read
\beq \label{set4}
M=\frac{\Omega_{(n+3)}}{16\pi G}\frac{\mu}{\Xi_1^2}\left( 2+\Xi_1(n+1)\right),\quad\quad J^{1}=\frac{\Omega_{(n+3)}}{8\pi G}\frac{\mu}{\Xi_1^2} a,\quad\quad J_{\perp}^{2}=\frac{\Omega_{(n+3)}}{8\pi G}\frac{\mu}{\Xi_1} b~,
\eeq
with
\beq\label{set5}
\Omega_{1}=\frac{1}{a},\quad\quad \Omega_{2}=\frac{b}{r_{+}^2+b^2}~,
\eeq
while the entropy and temperature are given by
\beq \label{set6}
S=\frac{\Omega_{(n+3)}}{4 G}\frac{a^2\mu}{\Xi_1 r_+},\quad\quad T=\frac{1}{4\pi r_{+}}\left( n - \frac{2b^2}{r_{+}^2+b^2} \right)~.
\eeq
Evaluating the conserved charges using Eqs.~\eqref{m}-\eqref{jt} results in
\beq
M=\frac{\Omega_{(n+3)}}{16\pi G}\frac{\tilde{r}_0^n(0)}{\Xi_1^2\Omega_{1}^2}\left( 2+\Xi_1(n+3)\right),\quad\quad J^{1}=\frac{\Omega_{(n+3)}}{8\pi G}\frac{\tilde{r}_0^n(0)}{\Xi_1\Omega_{1}^2}\frac{1}{\Omega_1},\quad\quad J_\perp^{2}=\frac{\Omega_{n+3)}}{8\pi G}\frac{\tilde{r}_0^n(0)}{\Xi_1\Omega_{1}^2}\tilde{b}(0)~.
\eeq
We can straightforwardly check that these results agree with the ones presented in \eqref{set4}-\eqref{set5} upon the identification $\tilde{r}_0^n(0)/\Omega_{1}^2=\mu$, $\Omega_{1}^{-1}=a_{1}$, and $\tilde{b}(0)=b$. Moreover, in the AdS case the tension (\ref{tension}) is non-zero and reads
\beq
\mathcal{T}=-\alpha\frac{\Omega_{(n+3)}}{8\pi G}\frac{\tilde{r}_0^n(0)}{\Xi_1^2\Omega_{1}^2}~.
\eeq
Using the Smarr relation \eqref{smarr} to compute the product $TS$ and then the first law of black hole thermodynamics we can exactly reproduce the temperature and entropy as given in \eqref{set6} in the same way we did for MP black holes. This leads one to conclude that higher-dimensional Kerr-(A)dS black holes in the ultra-spinning regime are correctly captured by the blackfold pole-dipole equations.

\addcontentsline{toc}{section}{References}

\bibliographystyle{newutphys}
\providecommand{\href}[2]{#2}\begingroup\raggedright\endgroup

\end{document}